\pdfoutput=1
\documentclass[12pt,a4paper]{article}
\usepackage{comment}
\usepackage{ifthen} % for conditional statements
\newboolean{pdflatex}
\setboolean{pdflatex}{true} % False for eps figures 

\newboolean{articletitles}
\setboolean{articletitles}{true} % False removes titles in references

\newboolean{uprightparticles}
\setboolean{uprightparticles}{false} %True for upright particle symbols

\def\paperauthors{LHCb collaboration} % Leave as is for PAPER, CONF and FIGURE
\def\paperasciititle{Measurement of X(3872) at 8 and 13 TeV} % Set ASCII title here
\def\papertitle{Measurement of $\theX$ production in proton-proton collisions at 
$\sqrt{s}=8$ and $13\tev$} % Latex formatted title
\def\paperkeywords{{High Energy Physics}, {LHCb}} % Comma separated list
\def\papercopyright{\the\year\ CERN for the benefit of the LHCb collaboration} % new since 9/Apr/2018
\def\paperlicence{CC BY 4.0 licence}
\def\paperlicenceurl{https://creativecommons.org/licenses/by/4.0/}

\def\theX     {{\ensuremath{\PX(3872)}}\xspace}
\def\jpsipipi     {{\ensuremath{\jpsi\pip\pim}}\xspace}
\def\XtoJpsipipi  {\decay{\theX}{\jpsi\pip\pim}}
\def\psitoJpsipipi  {\decay{\psi(2S)}{\jpsi\pip\pim}}

\def\effratio{\ensuremath{\epsilon_{\psitwos}/\epsilon_{\theX}}\xspace}

\def\yratio{\ensuremath{N_{\theX}/N_{\psitwos}}\xspace}
\def\fratio{\ensuremath{F_b^{\theX}/F_b^{\psitwos}}\xspace}

\usepackage[top=1in, bottom=1.25in, left=1in, right=1in]{geometry}

\usepackage{microtype}
\usepackage{lineno}  % for line numbering during review
\usepackage{xspace} % To avoid problems with missing or double spaces after
\usepackage{caption} %these three command get the figure and table captions automatically small

\usepackage{graphicx}  % to include figures (can also use other packages)
\usepackage{color}
\usepackage{colortbl}
\graphicspath{{./figs/}} % Make Latex search fig subdir for figures
\usepackage{amsmath} % Adds a large collection of math symbols
\usepackage{amssymb}
\usepackage{amsfonts}
\usepackage{upgreek} % Adds in support for greek letters in roman typeset

\newcommand*\patchAmsMathEnvironmentForLineno[1]{%
\expandafter\let\csname old#1\expandafter\endcsname\csname #1\endcsname
\expandafter\let\csname oldend#1\expandafter\endcsname\csname
end#1\endcsname
 \renewenvironment{#1}%
   {\linenomath\csname old#1\endcsname}%
   {\csname oldend#1\endcsname\endlinenomath}%
}
\newcommand*\patchBothAmsMathEnvironmentsForLineno[1]{%
  \patchAmsMathEnvironmentForLineno{#1}%
  \patchAmsMathEnvironmentForLineno{#1*}%
}
\AtBeginDocument{%
\patchBothAmsMathEnvironmentsForLineno{equation}%
\patchBothAmsMathEnvironmentsForLineno{align}%
\patchBothAmsMathEnvironmentsForLineno{flalign}%
\patchBothAmsMathEnvironmentsForLineno{alignat}%
\patchBothAmsMathEnvironmentsForLineno{gather}%
\patchBothAmsMathEnvironmentsForLineno{multline}%
\patchBothAmsMathEnvironmentsForLineno{eqnarray}%
}

\usepackage{hyperxmp}

\usepackage[pdftex,
            pdfauthor={\paperauthors},
            pdftitle={\paperasciititle},
            pdfkeywords={\paperkeywords},
            pdfcopyright={Copyright (C) \papercopyright},
            pdflicenseurl={\paperlicenceurl}]{hyperref}
\usepackage[bottom,flushmargin,hang,multiple]{footmisc}

\usepackage[all]{hypcap} % Internal hyperlinks to floats.

\usepackage{xspace} 
\usepackage{upgreek}

\def\lhcb   {\mbox{LHCb}\xspace}

\def\MagUp {\mbox{\em Mag\kern -0.05em Up}\xspace}

\ifthenelse{\boolean{uprightparticles}}%
{

 \def\Pmu         {\ensuremath{\upmu}\xspace}

 \def\Ppi         {\ensuremath{\uppi}\xspace}

 \def\Pchi        {\ensuremath{\upchi}\xspace}                 
 \def\Ppsi        {\ensuremath{\uppsi}\xspace}

 \def\PDelta      {\ensuremath{\Delta}\xspace}                 
 \def\PXi         {\ensuremath{\Xi}\xspace}                 
 \def\PLambda     {\ensuremath{\Lambda}\xspace}                 
 \def\PSigma      {\ensuremath{\Sigma}\xspace}                 
 \def\POmega      {\ensuremath{\Omega}\xspace}                 
 \def\PUpsilon    {\ensuremath{\Upsilon}\xspace}

 \def\PB      {\ensuremath{\mathrm{B}}\xspace}                 
                  
 \def\PD      {\ensuremath{\mathrm{D}}\xspace}

 \def\PJ      {\ensuremath{\mathrm{J}}\xspace}                 
 \def\PK      {\ensuremath{\mathrm{K}}\xspace}

 \def\Pb      {\ensuremath{\mathrm{b}}\xspace}                 
 \def\Pc      {\ensuremath{\mathrm{c}}\xspace}

 \def\Pi      {\ensuremath{\mathrm{i}}\xspace}

 \def\Ps      {\ensuremath{\mathrm{s}}\xspace}

 \def\thebaroffset{0.0em}
}
{

 \def\Pmu         {\ensuremath{\mu}\xspace}

 \def\Ppi         {\ensuremath{\pi}\xspace}

 \def\Pchi        {\ensuremath{\chi}\xspace}                 
 \def\Ppsi        {\ensuremath{\psi}\xspace}                 
                  
 \mathchardef\PDelta="7101
 \mathchardef\PXi="7104
 \mathchardef\PLambda="7103
 \mathchardef\PSigma="7106
 \mathchardef\POmega="710A
 \mathchardef\PUpsilon="7107
                  
 \def\PB      {\ensuremath{B}\xspace}                 
                  
 \def\PD      {\ensuremath{D}\xspace}

 \def\PJ      {\ensuremath{J}\xspace}                 
 \def\PK      {\ensuremath{K}\xspace}

 \def\Pb      {\ensuremath{b}\xspace}                 
 \def\Pc      {\ensuremath{c}\xspace}

 \def\Pi      {\ensuremath{i}\xspace}

 \def\Ps      {\ensuremath{s}\xspace}

 \def\thebaroffset{0.18em}
}
\newcommand{\offsetoverline}[2][\thebaroffset]{\kern #1\overline{\kern -#1 #2}}%

\makeatletter
\ifcase \@ptsize \relax% 10pt
  \newcommand{\miniscule}{\@setfontsize\miniscule{4}{5}}% \tiny: 5/6
\or% 11pt
  \newcommand{\miniscule}{\@setfontsize\miniscule{5}{6}}% \tiny: 6/7
\or% 12pt
  \newcommand{\miniscule}{\@setfontsize\miniscule{5}{6}}% \tiny: 6/7
\fi
\makeatother

\DeclareRobustCommand{\optbar}[1]{\shortstack{{\miniscule (\rule[.5ex]{1.25em}{.18mm})}
  \\ [-.7ex] $#1$}}

\def\mup        {{\ensuremath{\Pmu^+}}\xspace}
\def\mun        {{\ensuremath{\Pmu^-}}\xspace} % muon negative (\mum is taken)

\def\mumu       {{\ensuremath{\Pmu^+\Pmu^-}}\xspace}

\def\squark    {{\ensuremath{\Ps}}\xspace}

\def\cquark    {{\ensuremath{\Pc}}\xspace}

\def\bquark    {{\ensuremath{\Pb}}\xspace}

\def\pion   {{\ensuremath{\Ppi}}\xspace}

\def\pip    {{\ensuremath{\pion^+}}\xspace}
\def\pim    {{\ensuremath{\pion^-}}\xspace}

\def\KorKbar {\kern \thebaroffset\optbar{\kern -\thebaroffset \PK}{}\xspace}

\def\Dbar    {{\ensuremath{\offsetoverline{\PD}}}\xspace}
\def\D       {{\ensuremath{\PD}}\xspace}

\def\DorDbar {\kern \thebaroffset\optbar{\kern -\thebaroffset \PD}\xspace}

\def\Dp      {{\ensuremath{\D^+}}\xspace}
\def\Dm      {{\ensuremath{\D^-}}\xspace}

\def\DpDm    {\ensuremath{\Dp {\kern -0.16em \Dm}}\xspace}
\def\Dstar   {{\ensuremath{\D^*}}\xspace}
\def\Dstarb  {{\ensuremath{\Dbar{}^*}}\xspace}

\def\Dstarzb {{\ensuremath{\Dbar{}^{*0}}}\xspace}

\def\B       {{\ensuremath{\PB}}\xspace}

\def\BorBbar {\kern \thebaroffset\optbar{\kern -\thebaroffset \PB}\xspace}

\def\Bd      {{\ensuremath{\B^0}}\xspace}

\def\BdorBdbar {\kern \thebaroffset\optbar{\kern -\thebaroffset \Bd}\xspace}

\def\Bs      {{\ensuremath{\B^0_\squark}}\xspace}

\def\BsorBsbar {\kern \thebaroffset\optbar{\kern -\thebaroffset \Bs}\xspace}

\def\jpsi     {{\ensuremath{{\PJ\mskip -3mu/\mskip -2mu\Ppsi}}}\xspace}
\def\psitwos  {{\ensuremath{\Ppsi{(2S)}}}\xspace}

\def\Y#1S{\ensuremath{\PUpsilon{(#1S)}}\xspace}

\def\theX     {{\ensuremath{\Pchi_{c1}(3872)}}\xspace}

\def\LorLbar     {\kern \thebaroffset\optbar{\kern -\thebaroffset \PLambda}\xspace}

\def\BF         {{\ensuremath{\mathcal{B}}}\xspace}

\newcommand{\decay}[2]{\ensuremath{#1\!\to #2}\xspace} 

\def\to                 {\ensuremath{\rightarrow}\xspace}

\def\AT#1     {\ensuremath{A_{\mathrm{T}}^{#1}}\xspace}           % 2

\def\C#1      {\ensuremath{\mathcal{C}_{#1}}\xspace}                       % 9
\def\Cp#1     {\ensuremath{\mathcal{C}_{#1}^{'}}\xspace}                    % 7
\def\Ceff#1   {\ensuremath{\mathcal{C}_{#1}^{\mathrm{(eff)}}}\xspace}        % 9  
\def\Cpeff#1  {\ensuremath{\mathcal{C}_{#1}^{'\mathrm{(eff)}}}\xspace}       % 7
\def\Ope#1    {\ensuremath{\mathcal{O}_{#1}}\xspace}                       % 2
\def\Opep#1   {\ensuremath{\mathcal{O}_{#1}^{'}}\xspace}                    % 7

\newcommand{\nospaceunit}[1]{\ensuremath{\text{#1}}}       
\newcommand{\aunit}[1]{\ensuremath{\text{\,#1}}}       
\newcommand{\tev}{\aunit{Te\kern -0.1em V}\xspace}
\newcommand{\gev}{\aunit{Ge\kern -0.1em V}\xspace}
\newcommand{\mev}{\aunit{Me\kern -0.1em V}\xspace}
\newcommand{\kev}{\aunit{ke\kern -0.1em V}\xspace}
\newcommand{\ev}{\aunit{e\kern -0.1em V}\xspace}
 
\newcommand{\mevc}{\ensuremath{\aunit{Me\kern -0.1em V\!/}c}\xspace}
\newcommand{\gevc}{\ensuremath{\aunit{Ge\kern -0.1em V\!/}c}\xspace}
\newcommand{\mevcc}{\ensuremath{\aunit{Me\kern -0.1em V\!/}c^2}\xspace}
\newcommand{\gevcc}{\ensuremath{\aunit{Ge\kern -0.1em V\!/}c^2}\xspace}
\def\mum  {\ensuremath{\,\upmu\nospaceunit{m}}\xspace}

\def\fb   {\ensuremath{\aunit{fb}}\xspace}
\def\invfb   {\ensuremath{\fb^{-1}}\xspace}

\def\ps   {\ensuremath{\aunit{ps}}\xspace}

\newcommand{\chisqndf}{\ensuremath{\chi^2/\mathrm{ndf}}\xspace}

\def\gsim{{~\raise.15em\hbox{$>$}\kern-.85em
          \lower.35em\hbox{$\sim$}~}\xspace}
\def\lsim{{~\raise.15em\hbox{$<$}\kern-.85em
          \lower.35em\hbox{$\sim$}~}\xspace}

\def\sPlot{\mbox{\em sPlot}\xspace}

\def\sqs   {\ensuremath{\protect\sqrt{s}}\xspace}
\newcommand{\sqrts}[1]{\ensuremath{\sqs = #1\,\tev}\xspace}

\def\pt         {\ensuremath{p_{\mathrm{T}}}\xspace}

\def\ptot       {\ensuremath{p}\xspace}

\def\evtgen     {\mbox{\textsc{EvtGen}}\xspace}

\def\geant      {\mbox{\textsc{Geant4}}\xspace}

\def\photos     {\mbox{\textsc{Photos}}\xspace}

\def\pythia     {\mbox{\textsc{Pythia}}\xspace}

\def\tell1  {TELL1\xspace}
\def\ukl1   {UKL1\xspace}

\usepackage{cite} % Allows for ranges in citations
\usepackage{mciteplus}
\usepackage{longtable} % only for template; not usually to be used in PAPERs

\usepackage{rotating}
\usepackage{multirow}

\usepackage{enumitem}

\RequirePackage[normalem]{ulem} %DIF PREAMBLE
\RequirePackage{color}\definecolor{RED}{rgb}{1,0,0}\definecolor{BLUE}{rgb}{0,0,1} %DIF PREAMBLE
\begin{document}

%%%%%%%%%%%%%%%%%%%%%%%%%
%%%%% Title     %%%%%%%%%
%%%%%%%%%%%%%%%%%%%%%%%%%
\renewcommand{\thefootnote}{\fnsymbol{footnote}}
\setcounter{footnote}{1}

% %%%%%%% CHOOSE TITLE PAGE--------
%\onecolumn
% $Id: title-LHCb-PAPER.tex 122889 2018-08-17 17:59:55Z pkoppenb $
% ===============================================================================
% Purpose: LHCb-PAPER journal paper title page template
% Author: 
% Created on: 2010-09-25
% ===============================================================================

%%%%%%%%%%%%%%%%%%%%%%%%%
%%%%%  TITLE PAGE  %%%%%%
%%%%%%%%%%%%%%%%%%%%%%%%%
\begin{titlepage}
\pagenumbering{roman}

% Header ---------------------------------------------------
\vspace*{-1.5cm}
\centerline{\large EUROPEAN ORGANIZATION FOR NUCLEAR RESEARCH (CERN)}
\vspace*{1.5cm}
\noindent
\begin{tabular*}{\linewidth}{lc@{\extracolsep{\fill}}r@{\extracolsep{0pt}}}
\ifthenelse{\boolean{pdflatex}}% Logo format choice
{\vspace*{-1.5cm}\mbox{\!\!\!\includegraphics[width=.14\textwidth]{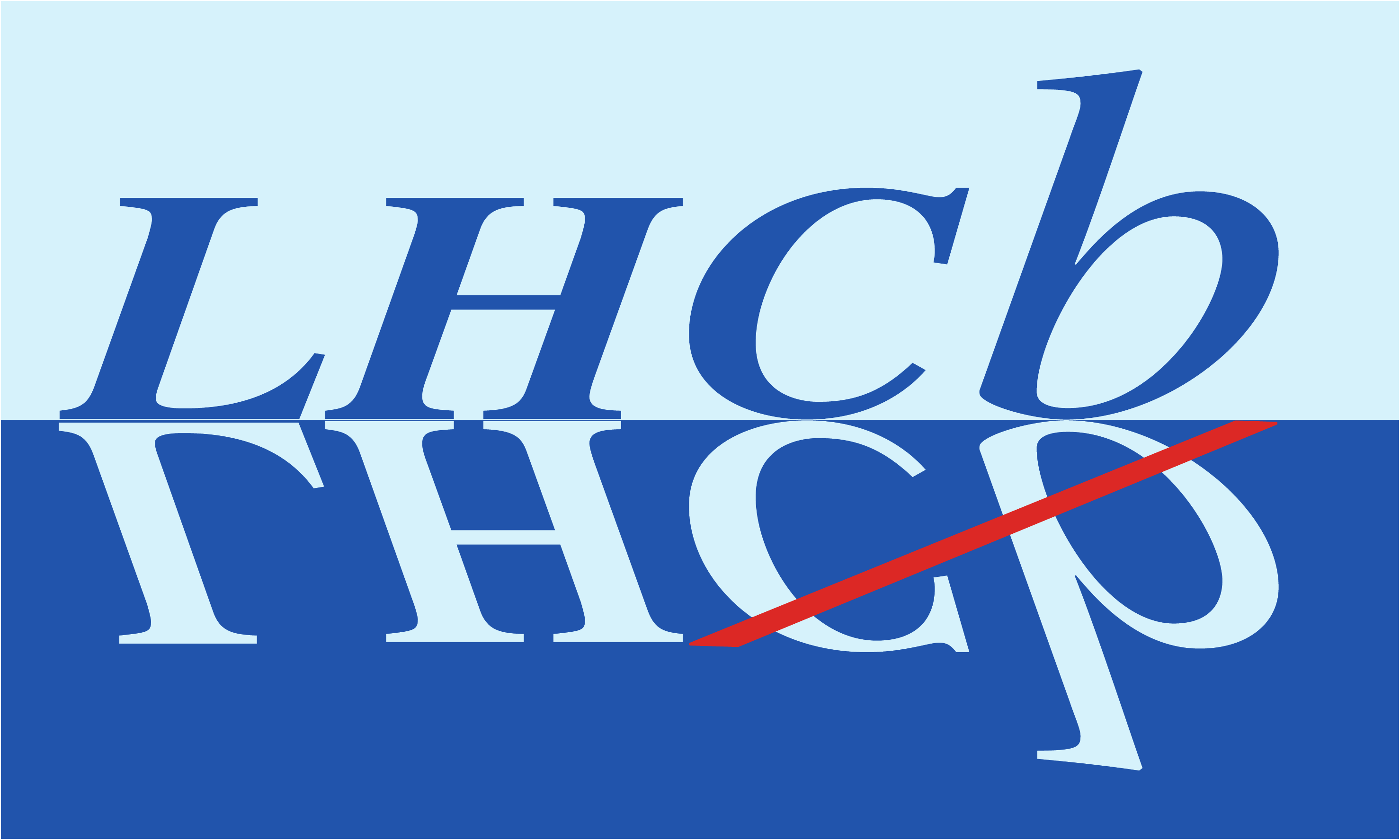}} & &}%
{\vspace*{-1.2cm}\mbox{\!\!\!\includegraphics[width=.12\textwidth]{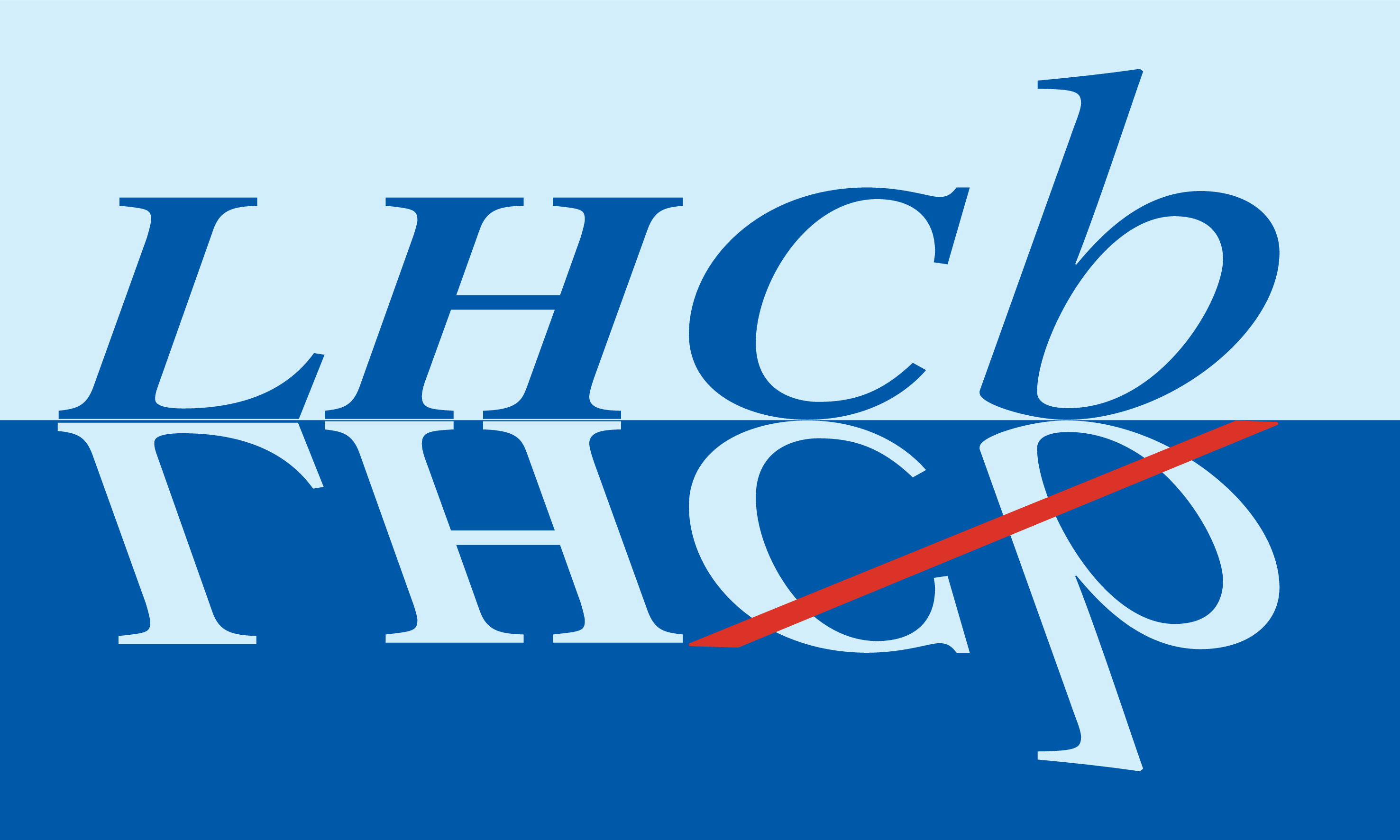}} & &}%
\\
 & & CERN-EP-2021-182 \\  % ID 
 & & LHCb-PAPER-2021-026 \\  % ID 
 %& & \today \\ % Date - Can also hardwire e.g.: 23 March 2010
 & & 28 January 2022 \\ % Date - Can also hardwire e.g.: 23 March 2010
 & & \\
% not in paper \hline
\end{tabular*}

\vspace*{1.0cm}

% Title --------------------------------------------------
{\normalfont\bfseries\boldmath\huge
\begin{center}
% DO NOT EDIT HERE. Instead edit macro in main.tex to keep metadata correct
  \papertitle 
\end{center}
}

\vspace*{1.0cm}

% Authors -------------------------------------------------
\begin{center}
%In the footnote, replace 'paper' by 'Letter' in case of submission to PRL or PLB 
% Edit macro in main.tex to keep metadata correct
\paperauthors\footnote{Authors are listed at the end of this paper.}
\end{center}

\vspace{\fill}

% Abstract -----------------------------------------------
\begin{abstract}
  \noindent
The production cross-section of the \theX state relative to the \psitwos meson is measured using proton-proton collision data collected with
the LHCb experiment at centre-of-mass energies of $\sqrt{s}=8$ and $13\tev$, corresponding to integrated luminosities of 2.0 and 5.4\invfb, respectively. The two mesons are reconstructed in the \jpsipipi final state. 
The ratios of the prompt and nonprompt \theX to \psitwos production cross-sections are measured as a function of transverse momentum, $\pt$, and rapidity, $y$, of the \theX and \psitwos states, 
in the kinematic range \mbox{$4 < \pt < 20$} \gevc and \mbox{$2.0 < y < 4.5$}. The prompt ratio is found to increase with \pt, independently of $y$. 
For the prompt component,
the double ratio of the \theX 
and \psitwos production cross-sections 
between 13 and 8\tev is observed to be consistent with unity, independent of \pt and centre-of-mass energy.

\end{abstract}

\vspace*{2.0cm}

\begin{center}
Published in JHEP 01 (2022) 131
\end{center}

\vspace{\fill}

{\footnotesize 
% Edit macro in main.tex to keep metadata correct
\centerline{\copyright~\papercopyright. \href{\paperlicenceurl}{\paperlicence}.}}
\vspace*{2mm}

\end{titlepage}

%%%%%%%%%%%%%%%%%%%%%%%%%%%%%%%%
%%%%%  EOD OF TITLE PAGE  %%%%%%
%%%%%%%%%%%%%%%%%%%%%%%%%%%%%%%%

%  empty page follows the title page ----
\newpage
\setcounter{page}{2}
\mbox{~}
%\newpage
%
%% Author List ----------------------------
%%  You need to get a new author list!
%\input{LHCb_authorlist.tex}
%
%The author list for journal publications is provided by the Membership Committee shortly after 'approval to go to paper' has been given.
%%It will be made available on the page
%%\verb!http://www.physik.uzh.ch/~strauman/forMemCo/LHCb-PAPER-XXXX-XXX/! .
%It will be sent to you by email shortly after a paper number has beens assigned.
%The author list should be included already at first circulation, 
%to allow new members of the collaboration to verify whether they have been included correctly.
%Occasionally a misspelled name is corrected or associated institutions become full members.
%In that case, a new author list will be sent to you.
%In case line numbering doesn't work well after including the authorlist, try moving the \verb!\bigskip! after the last author to a separate line.
%
%
%The authorship for Conference Reports should be ``The LHCb
%  collaboration'', with a footnote giving the name(s) of the contact
%  author(s), but without the full list of collaboration names.

\cleardoublepage

%\twocolumn
% %%%%%%%%%%%%% ---------

\renewcommand{\thefootnote}{\arabic{footnote}}
\setcounter{footnote}{0}

%%%%%%%%%%%%%%%%%%%%%%%%%%%%%%%%
%%%%%  Table of Content   %%%%%%
%%%%%%%%%%%%%%%%%%%%%%%%%%%%%%%%
%%%% Uncomment next 2 lines if desired
%\tableofcontents
%\cleardoublepage

%%%%%%%%%%%%%%%%%%%%%%%%%
%%%%% Main text %%%%%%%%%
%%%%%%%%%%%%%%%%%%%%%%%%%

\pagestyle{plain} % restore page numbers for the main text
\setcounter{page}{1}
\pagenumbering{arabic}

%\linenumbers

\section{Introduction}
\label{sec:Introduction}

The \theX state was observed in the \jpsipipi invariant-mass spectrum by the Belle collaboration in 2003~\cite{Belle_X3872obs}, and was subsequently 
confirmed by the BaBar, CDF and D0 collaborations~\cite{BABAR_X3872obs,CDF_X3872obs,D0_X3872obs}.
Its quantum numbers have been determined to be $J^{PC}=$
$1^{++}$ by the LHCb collaboration~\cite{LHCb-PAPER-2013-001}.
Nevertheless, despite intense experimental and theoretical studies, the nature of the state is still unclear.
The mass is close to the $D^0\Dstarzb$  threshold, which led to
models where the \theX state is a $D^0\Dstarzb$ molecule with a very small
binding energy~\cite{X3872asDD,X3872asDD2}.
The LHCb collaboration indeed measured that the mass is slightly below that threshold ~\cite{LHCb-PAPER-2020-008,LHCb-PAPER-2020-009}.
However, the differential production cross-section measured by the CMS
collaboration~\cite{Chatrchyan:2013cld} is lower than that predicted by non-relativistic QCD (NRQCD)~\cite{Artoisenet:2009wk} for the
$D^0\Dstarzb$ molecule hypothesis. 
If the \theX state were a weakly bound molecule state, it could also be produced 
by the creation of a $\Dstar\Dstarb$ pair at short distance followed by a rescattering into the $\theX\pi$ final state~\cite{Braaten:2019sxh,Braaten:2019yua}. However the production of the \theX accompanied by a pion has not been observed so far~\cite{D0:2020nce}.
Alternatively, the \theX state can be interpreted as an admixture of $\chi_{c1}(2P)$ and $D^0\Dstarzb$ molecule states, 
produced through its $\chi_{c1}(2P)$ component~\cite{Meng:2005er}. 
Under this hypothesis, a next-to-leading-order (NLO) NRQCD calculation~\cite{Meng:2013gga} tuned to the results obtained by the CMS collaboration,  agrees well with measurements performed by the ATLAS collaboration in a much wider range of the $\theX$ transverse momentum~\cite{Aaboud:2016vzw}. 
Recently, the ratio of prompt-production cross-sections between \theX and \psitwos states produced directly from proton-proton ($pp$) collisions,
as a function of multiplicity of the charged particles in an event,
has been measured by the LHCb collaboration using 8\tev $pp$ collision data~\cite{LHCb-PAPER-2020-023}.  
This ratio is found to decrease with multiplicity.
The interpretation 
of this observation is still unclear~\cite{Esposito:2020ywk, Braaten:2020iqw}.
Recent measurements of the $\Bs\rightarrow\theX\phi$ branching fraction~\cite{CMS:2020eiw,LHCb-PAPER-2020-035} would support the interpretation of the \theX state as a tetraquark~\cite{Maiani:2020zhr}.

In this paper, the double-differential production cross-section of the \theX state relative to that of the \psitwos meson, where both decay to \jpsipipi with \jpsi decaying to \mumu final state, is measured 
using $pp$ collision data collected by the LHCb detector at centre-of-mass energies of $\sqrt{s}=8$ and $13\tev$.  
The cross-section is determined in intervals of the \jpsipipi transverse momentum, \pt, and rapidity, $y$, within the ranges \mbox{$4<\pt<20\gevc$} and \mbox{$2.0<y<4.5$}.
The cross-section ratio 
$\sigma_{\theX}/\sigma_{\psitwos}$
is measured separately for prompt and nonprompt production of the \theX and \psitwos mesons, the latter occurring via $b$-hadron decays.
In this ratio, the systematic uncertainties largely cancel. 
The production cross-section of the \theX state at a centre-of-mass energy of \mbox{\sqrts{7}}
has been previously measured with 35 pb$^{-1}$ of $pp$ collision data~\cite{LHCb-PAPER-2011-034}.
Using the data recorded during the 2012 and 2016--2018 data-taking periods, corresponding to integrated luminosities of 
2.0 and 5.4\invfb, 
the signal yields increase by about a factor of 400,
allowing a measurement of the double-differential cross-section to be performed for the first time.

\section{Detector and simulation}
\label{sec:Detector}

The \lhcb detector~\cite{LHCb-DP-2008-001,LHCb-DP-2014-002} is a single-arm forward spectrometer covering the \mbox{pseudorapidity} range $2<\eta <5$,
designed for the study of particles containing \bquark or \cquark quarks. The detector includes a high-precision tracking system
consisting of a silicon-strip vertex detector surrounding the $pp$
interaction region~\cite{LHCb-DP-2014-001},
a large-area silicon-strip detector located
upstream of a dipole magnet with a bending power of about
$4{\mathrm{\,Tm}}$, and three stations of silicon-strip detectors and straw
drift tubes~\cite{LHCb-DP-2013-003,LHCb-DP-2017-001}
placed downstream of the magnet.
The tracking system provides a measurement of the momentum, \ptot, of charged particles with
a relative uncertainty that varies from 0.5\% at low momentum to 1.0\% at 200\gevc.
The minimum distance of a track to a primary $pp$-collision vertex (PV), the impact parameter (IP), 
is measured with a resolution of $(15+29/\pt)\mum$, with \pt, in\,\gevc.
Different types of charged hadrons are distinguished using information
from two ring-imaging Cherenkov (RICH) detectors\cite{LHCb-DP-2012-003}. 
The online event selection is performed by a trigger system~\cite{LHCb-DP-2012-004}, 
which consists of a hardware stage, based on information from the muon
system, followed by a software stage, which is applied to perform a full event reconstruction.
To avoid domination of the trigger CPU time by a few events with high occupancy, a set of global event requirements\cite{LHCb-DP-2012-004} is applied on the hit multiplicity of each sub-detector used by the pattern recognition algorithms. These requirements reject high-multiplicity events with a large number of $pp$ interactions.

Simulated samples are used to develop the event selection and
to estimate the detector acceptance and the
efficiency of the imposed selection requirements.
Simulated $pp$ collisions are generated using
\pythia~\cite{Sjostrand:2007gs,Sjostrand:2006za} 
with a specific \lhcb configuration~\cite{LHCb-PROC-2010-056}.  
Decays of unstable particles
are described by \evtgen~\cite{Lange:2001uf}, in which final-state
radiation is generated using \photos~\cite{davidson2015photos}. The
interaction of the generated particles with the detector, and its response,
are implemented using the \geant
toolkit~\cite{Allison:2006ve, Agostinelli:2002hh} as described in
Ref.~\cite{LHCb-PROC-2011-006}.

\section{Event selection}
\label{sec:evt_sel}
In the online event selection, 
signal candidates are required to pass 
dedicated $\jpsi$ trigger algorithms. 
These algorithms require at least one muon to have high transverse momentum at the hardware stage, 
and a pair of oppositely-charged muon candidates to originate from a common vertex and to have an invariant mass in a wide window around the $\jpsi$ mass at the software stage.

The $\theX$ and $\psitwos$ candidates are both reconstructed in the 
$\jpsi\pi^{+}\pi^{-}$ final state, with the \jpsi meson decaying into a pair of oppositely charged muons.
At least one reconstructed primary vertex is required per event.
Muon candidates are required to be well identified, and have \mbox{$\pt > 650\mevc$} and \mbox{$p > 10\gevc$}.
Only reconstructed muon tracks of good quality are selected.
The \mumu pair is required to have a combined $\pt > 3 \gevc$ and an invariant mass in the range 3010--3170 \mevcc.
The \chisqndf of the dimuon
vertex fit is required to be less than 20, where ndf is the number of degrees of freedom.

Charged pion candidates are selected using particle identification (PID) information from the RICH detectors.
The (transverse) momentum of the pions is required to be greater than 
(500) 3000\mevc, while the pion track \chisqndf to  
be less than 4. 

The \theX and \psitwos candidates are reconstructed by combining
each \jpsi candidate with a pair of oppositely charged pions. In order to improve the $\jpsi\pi^{+}\pi^{-}$ invariant-mass resolution, a vertex fit constraining the \mumu invariant mass to the known 
\jpsi mass~\cite{PDG2020}, 
$m_{\jpsi}=3096.9 \mevcc$, is performed. The vertex fit \chisqndf is required to be less than 5. The decay energy release, $Q\equiv  M_{\jpsipipi} - m_{\jpsi} -
M_{\pi^{+}\pi^{-}}$, 
with $M_{\jpsipipi}$ and $M_{\pi^{+}\pi^{-}}$ being the invariant masses of, respectively, 
\jpsipipi and ${\pi^{+}\pi^{-}}$ systems, 
is required to be less than 300\mevcc.  

A pseudo-decay-time of the \theX and \psitwos candidates is constructed as
  \begin{equation}
    \label{eq:tz_fun}
t_z = \frac{(z - z_{\rm PV})\times m}{p_z},
  \end{equation}
where $z$ and $z_{\rm PV}$ are the candidate decay vertex and best-reconstructed PV positions along the beam ($z$) axis, $p_z$ is the $z$ component of the \theX or \psitwos momenta, and $m$ represents the known masses of these states~\cite{PDG2020}. Only candidates with $|t_z| < 10\ps$ are kept for further analysis.
The pseudo-decay-time of promptly produced \theX and \psitwos mesons is zero, whereas that of mesons originating from $b$-hadron decays follows an exponential distribution. This variable is used to statistically  discriminate between promptly and nonpromptly 
produced candidates.

\section{Cross-section determination}
\label{sec:cs_determination}

The differential production cross-section of \theX relative to \psitwos mesons times their ratio of branching fractions (\BF) to the \jpsipipi final state measured in $(\pt, y)$ intervals 
is defined as
  \begin{equation}
    \label{eq:cs_ratio_fun}
R \equiv \frac{\sigma_{\theX}}{\sigma_{\psitwos}}\times\frac{\BF(\theX\to\jpsi\pip\pim)}{\BF(\psitoJpsipipi)}  = \frac{N_{\theX}}{N_{\psitwos}}\times\frac{\epsilon_{\psitwos}}{\epsilon_{\theX}},
  \end{equation}
where $N_{\theX}$ and $N_{\psitwos}$ are the observed signal yields of $\theX$ and $\psitwos$ mesons, and  
$\epsilon_{\theX}$  and $\epsilon_{\psitwos}$ are the total efficiencies, respectively.
The yields of prompt \theX and \psitwos mesons and those from $b$-hadron decays are determined in each $(\pt,y)$ interval from a two-dimensional
extended binned maximum-likelihood fit to the $\jpsipipi$ invariant-mass spectrum and the pseudo-decay-time distribution. 
As the trigger thresholds varied for different periods of data taking, the signal yields and efficiencies are determined separately for each year. 

For the invariant-mass model, the sum of two double-sided Crystal Ball ($F_{\rm DSCB}$) functions~\cite{DESY-F31-86-02} is used to describe the $\psitwos$ signal. The two $F_{\rm DSCB}$ functions share a common mean ($\mu$) and have different width parameters $\sigma_{1}^{\psitwos}$ and $\sigma_{2}^{\psitwos}$. The relative fraction ($f$) of the $F_{\rm DSCB}$ functions is determined from simulation.
The radiative-tail parameters of the $F_{\rm DSCB}$ functions ($\alpha_l,\alpha_r,n$) are parameterised as a function of the mass resolution, which is obtained using simulated samples.
The combinatorial background in the \psitwos signal window,
defined by $M_{\jpsipipi}$ within the 3650--3720\mevcc interval, is described by an exponential function ($F_{\rm Exp}$) with the slope parameter $(c_0)$ freely varied.
The invariant-mass model for the selected $\psitwos$ candidates can be written as
  \begin{align}
    \label{eq:psi_fun}
 F_{\psitwos}(m) & =  N_{\psitwos}(f\cdot F_{\rm DSCB}(m|\mu,\sigma_{1}^{\psitwos},\alpha_l,\alpha_r,n) \nonumber\\[1mm]
 & \qquad\qquad  + (1-f)\cdot F_{\rm DSCB}(m|\mu,\sigma_{2}^{\psitwos},\alpha_l,\alpha_r,n)) \nonumber\\[1mm]
& + N_{\rm bkg}^{\psitwos} F_{\rm Exp}(m|c_0),
  \end{align}
where $N_{\psitwos}$ is the signal yield of the $\psitwos$ meson, and $N_{\rm bkg}^{\psitwos}$ is the number of background events.

The fit function for the $\theX$ signal is defined with a non-relativistic Breit--Wigner shape ($F_{\rm BW}$) convolved with an invariant-mass resolution function, which has 
the same parameterisation as the \psitwos signal.
The worse signal-to-background ratio 
in the \theX mass region is mainly due to the fact that the production rate of \theX state is much smaller than that of the \psitwos charmonium.
The \theX mass ($M$) is fixed to the known \psitwos mass~\cite{PDG2020}, $m_{\psitwos}=3686.10\mevcc$, 
shifted by 185.49\mevcc~\cite{LHCb-PAPER-2020-009}, 
and its width ($\Gamma$) is Gaussian constrained to \mbox{1.19 $\pm$ 0.19 MeV},
which is the average value of two previous LHCb measurements~\cite{LHCb-PAPER-2020-009, LHCb-PAPER-2020-008}.
The ratios of the mass resolutions between the \theX and \psitwos signals are determined from simulated samples. 
An exponential function is used to describe the combinatorial background in the $\theX$ mass window, with $M_{\jpsipipi}$ between 3830 and 3910\mevcc.
The invariant-mass model for the selected $\theX$ candidates can be written as
\begin{align}
    \label{eq:x_fun}
 F_{\theX}(m)&=
     N_{\theX}(f\cdot F_{\rm BW}(m|M,\Gamma)\otimes F_{\rm DSCB}(m|\mu,\sigma_{1}^{\theX},\alpha_l,\alpha_r,n) \nonumber\\[1mm]
     &\qquad\qquad\quad +(1-f)\cdot F_{\rm BW}(m|M,\Gamma)\otimes F_{\rm DSCB}(m|\mu,\sigma_{2}^{\theX},\alpha_l,\alpha_r,n)) \nonumber\\[1mm]
     &+N_{\rm bkg} F_{\rm Exp}(m|c_0),
\end{align}
where $N_{\theX}$ is the signal yield of the $\theX$ state, and $N_{\rm bkg}$ is the number of background events. 

For the pseudo-decay-time model, a delta function is used to describe the $t_z$ distribution of the
prompt $\theX$ and $\psitwos$ signals, while an exponential function is used for that from $b$-hadron decays.
Both are convolved  with a resolution function ($F_\mathrm{resolution}$) chosen to be the sum of three Gaussian functions.
The average peseudodecay-time of the nonprompt \psitwos signal, referred to hereafter as pseudo-lifetime ($\tau_b^\prime$), is allowed to vary freely in the fit and found to be around 1.5\ps with a mild dependence on the \psitwos kinematics.
Due to the high level of background in the \theX candidate sample, the pseudo-lifetime of nonprompt \theX candidates is fixed to 1.5\ps. 
It is possible that the reconstructed \theX or \psitwos candidate is associated to a wrong PV,
which would result in a long tail in the $t_z$ distribution and would weakly contribute to the signal peak in the mass distribution. A nonparametric model is defined for this component by combining each signal candidate with the closest PV in a different event of the selected sample and taking the resulting $t_z$ distribution as a template in the fit.
The dominant background component is combinatorial, in which the \jpsi candidate is combined with a random pion pair uncorrelated with the signal candidate.
The pseudo-decay-time model of the \theX and \psitwos states can be written as 
\begin{align}
F_{t_z}(t_z)
      =&
         \left(N_{\mathrm{prompt}}\delta(t_z)+\frac{N_{\mathrm{nonprompt}}}{\tau_b^\prime}e^{-t_z/\tau_b^\prime}\right) \otimes F_\mathrm{resolution}(t_z) \nonumber \\
      &+N_{\mathrm{tail}} F_\mathrm{tail}(t_z)+N_\mathrm{bkg} F_\mathrm{background}(t_z), \label{eq:tz_fit}
\end{align}
where $N_{\mathrm{prompt}}$, $N_{\mathrm{nonprompt}}$,  $N_{\mathrm{tail}}$ and $N_\mathrm{bkg}$ are the number of prompt \theX(\psitwos) states, \theX(\psitwos) from $\bquark$-hadron decay, the wrong-PV candidates and background event yields, respectively.

\begin{figure}[tb]
  \begin{center}
    \includegraphics[width=0.49\linewidth]{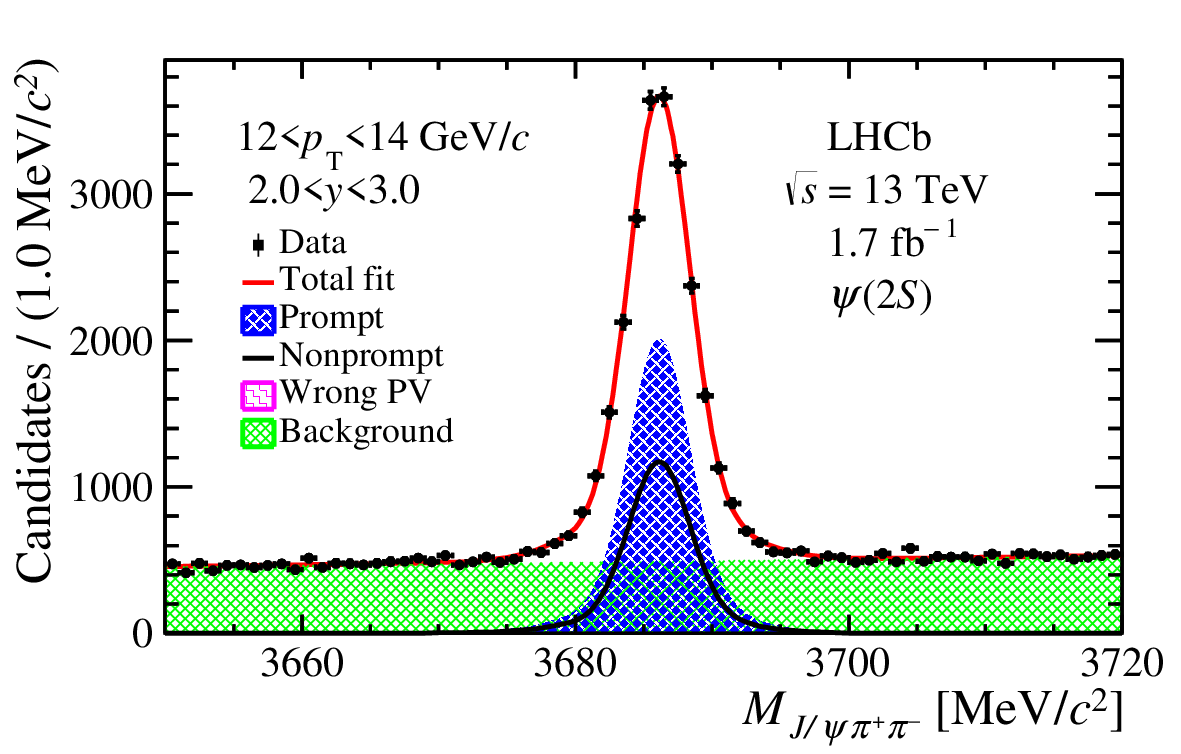}  
    \includegraphics[width=0.49\linewidth]{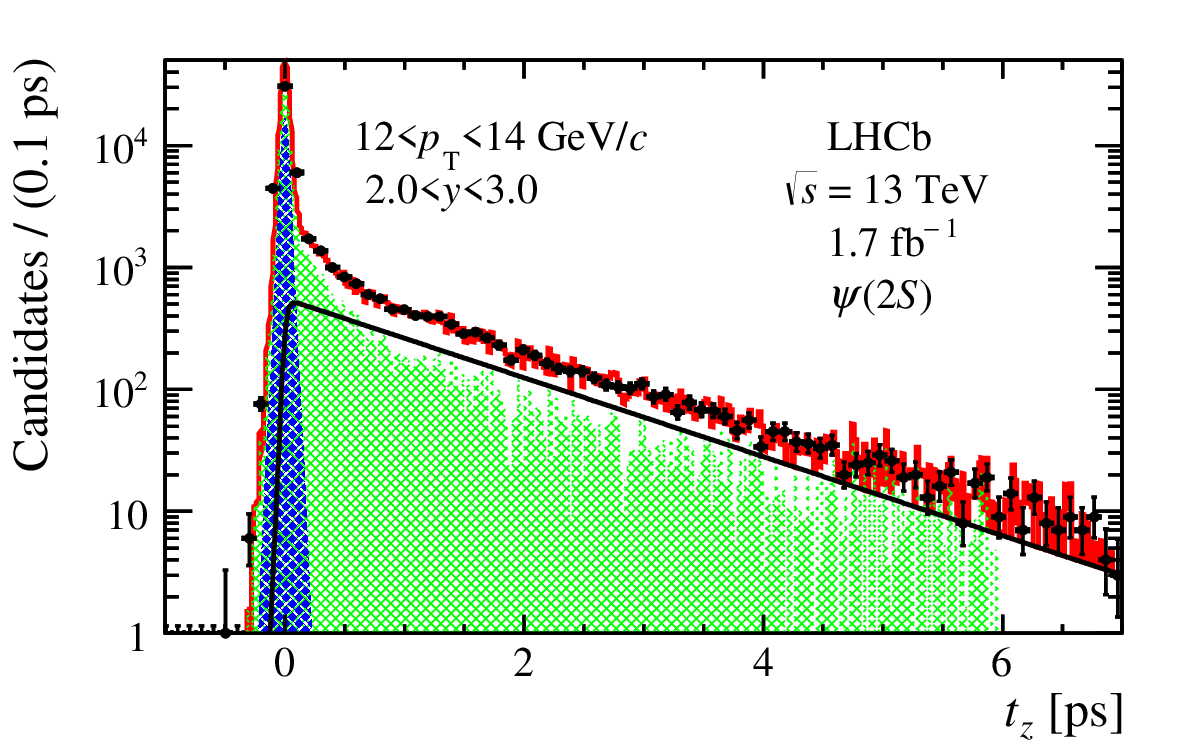} 

    \includegraphics[width=0.49\linewidth]{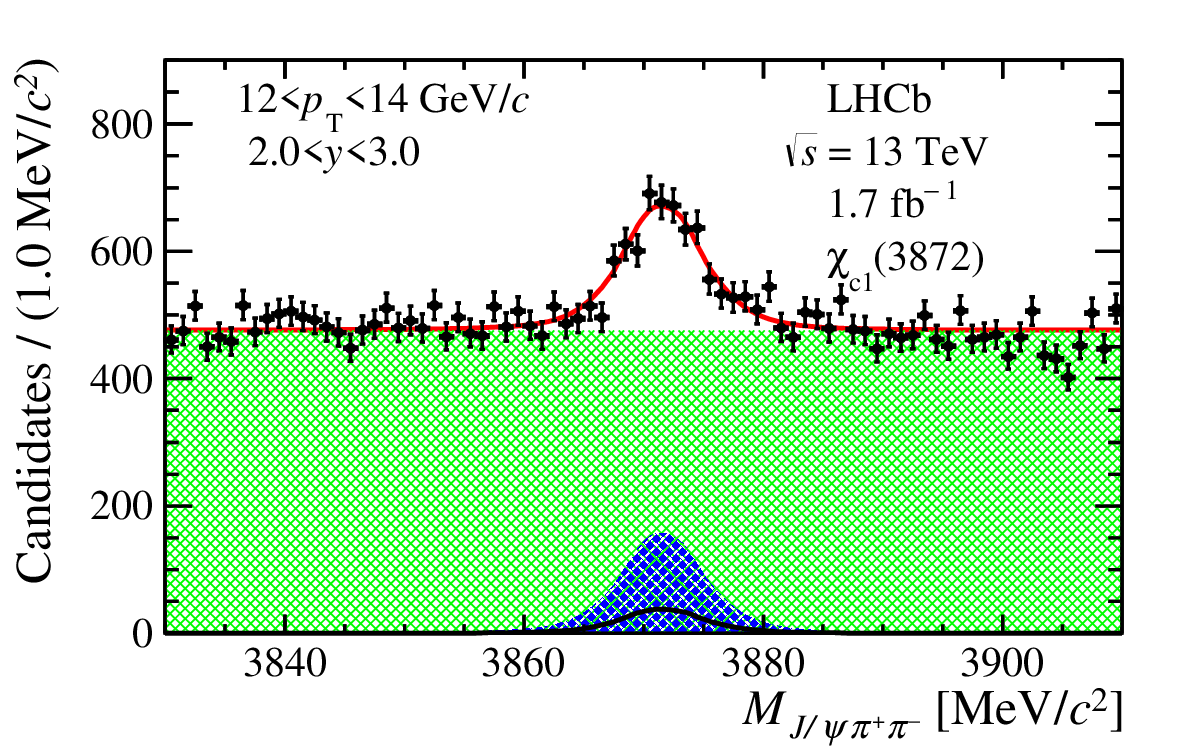}    
    \includegraphics[width=0.49\linewidth]{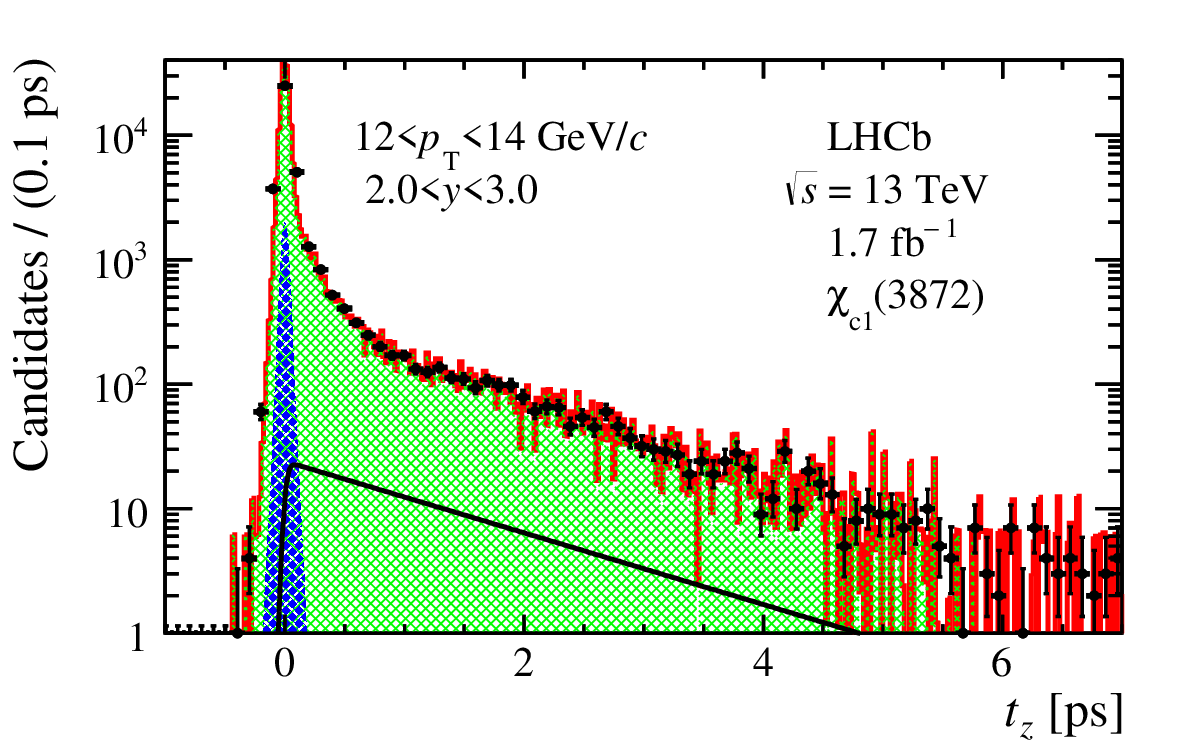}   
    \vspace*{-0.5cm}
  \end{center}
  \caption{Distributions of (left) invariant mass and (right) pseudo-decay-time for (top) \psitwos and (bottom) \theX candidates in the kinematic interval $12<\pt<14\gevc$ and $2.0<y<3.0$ for the 2016 data sample. Fit projections are overlaid. 
The solid red curve represents the total fit projection and the shaded green area corresponds to the background component. The prompt contribution of \theX and \psitwos mesons is shown
as the cross-hatched blue area, whereas the corresponding nonprompt component from \bquark-hadron decays is illustrated as a solid black line. The wrong PV contribution is consistent with zero.
}
  \label{fig:2d_fit}
\end{figure}

The $t_z$ distribution of the background in each $(\pt, y)$ interval is obtained using the \sPlot technique~\cite{Pivk:2004ty} with the \jpsipipi invariant mass as a discriminating observable.
The resulting model is fixed in the combined invariant mass and pseudo-decay-time fit. The fit is performed separately for each data-taking year.
As an example, Fig.~\ref{fig:2d_fit} shows the $M_{\jpsipipi}$ and $t_z$ distributions along with the fit projections for the 2016 data sample in the kinematic interval $12<\pt<14\gevc$ and $2<y<3$. 

\begin{figure}[!tb]
  \begin{center}
    \includegraphics[width=0.49\linewidth]{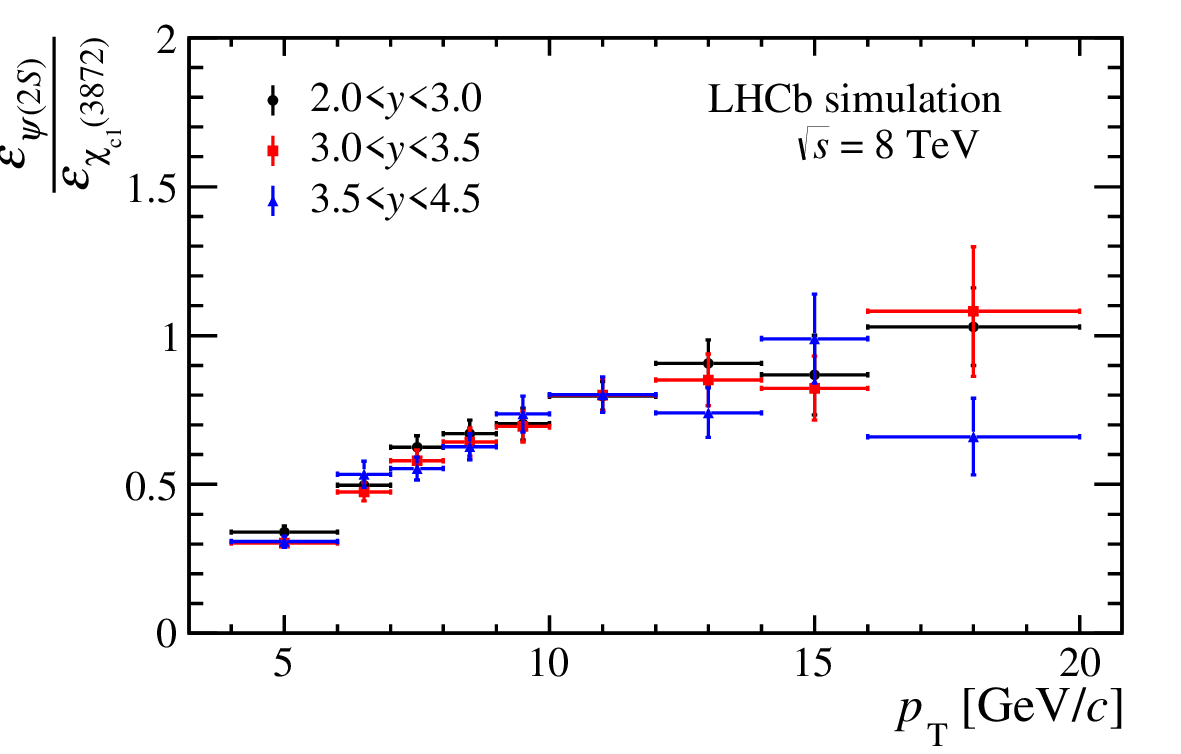} 
    \includegraphics[width=0.49\linewidth]{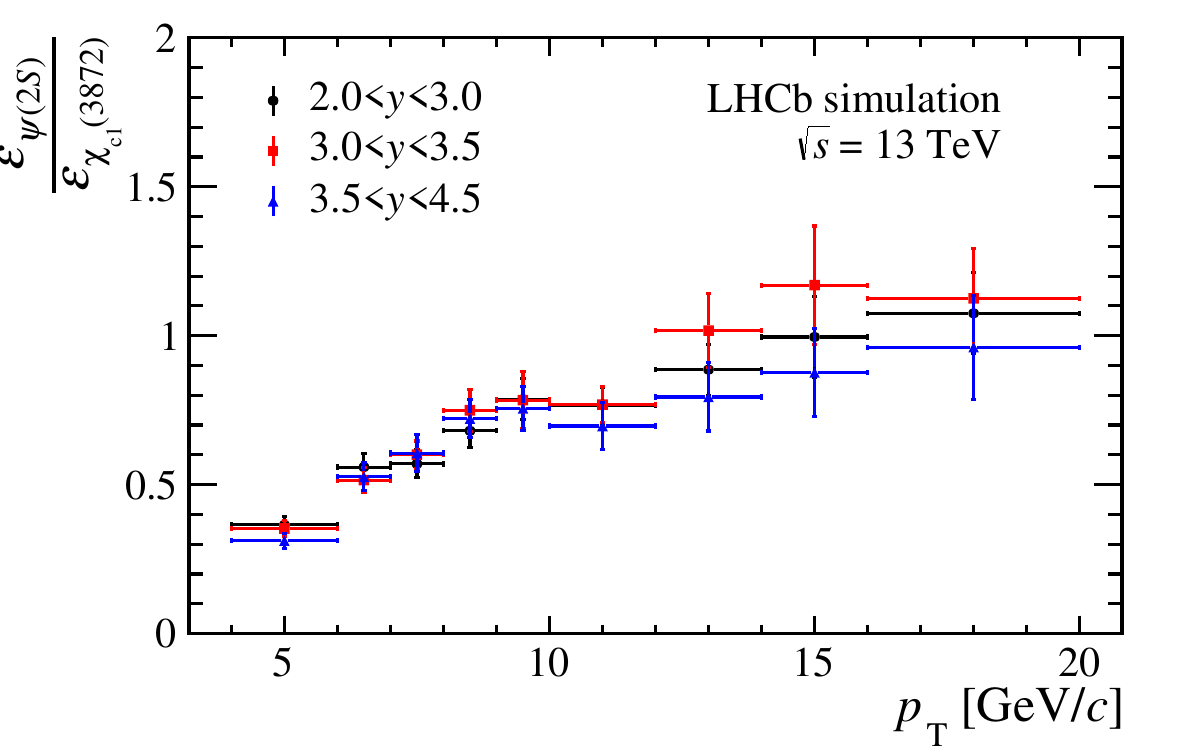} 
    \vspace*{-0.5cm}
  \end{center}
  \caption{
Efficiency ratio of \psitwos to \theX mesons as a function of \pt in three rapidity intervals.
    }
  \label{fig:eff_ratio_fig}
\end{figure}

The total efficiency in each kinematic interval is determined as the product of detector geometrical acceptance, particle reconstruction, event selection including trigger requirements, and particle identification efficiencies.  
The geometrical acceptance is calculated separately from \theX and \psitwos simulated  events. 
The track reconstruction and the particle-identification efficiencies are evaluated using simulated samples calibrated with data.
The efficiencies for prompt and nonprompt \theX and \psitwos signals are found to be slightly different, which is 
mainly caused by events containing \bquark-hadron decays having larger occupancies and thus smaller tracking efficiencies. 
This effect does not affect the ratio of \psitwos and \theX efficiencies, which is present in the cross-section ratio in Eq~\ref{eq:cs_ratio_fun}. 
The ratio of the total efficiencies of \psitwos to \theX mesons is shown in Fig.~\ref{fig:eff_ratio_fig} for the 2012 and 2016 data taking periods. The smaller efficiency at low \pt of the $\psitwos$ meson is due to its smaller mass. 

\section{Systematic uncertainties}
\label{sec:sys_uncer}
A variety of systematic uncertainty sources is studied and summarised in Table~\ref{tab:sys_sum}. 
The uncertainties arise from the \jpsipipi invariant mass and pseudo-decay-time fit models, and the computation of efficiencies. 
Some uncertainties depend on kinematics, with the largest values always appearing in the intervals with smaller sample sizes.

The signal lineshape chosen in the invariant-mass model can affect the measured signal yields.
Such effects are evaluated using pseudoexperiments
in which the signal description is taken from the simulated sample, and the background is generated with the shape and fraction determined from the fits to the data. 
The same fit model as used for the data is applied to these samples.
The difference between the fitted value of the \yratio ratio and the input value is taken as systematic uncertainty.
In the default fit, the parameters of the fit model, such as the fraction of two $F_{\rm DSCB}$ functions, the resolution ratios $\sigma_{2}^{\psitwos}/\sigma_{1}^{\psitwos}$ and $\sigma_{i}^{\theX}/\sigma_{i}^{\psitwos}\, (i=1,2)$ are fixed in the fit to the data.
These parameters are varied within their statistical uncertainties, and the average shift of the \yratio ratio is assigned as uncertainty. 
The contributions to the systematic uncertainty due to the background mass shape are estimated by replacing the exponential function
by a second-order polynomial function, and evaluating the difference of \yratio between the alternative and the default fits. 

There are several analysis choices that could affect the nonprompt fit fraction, $F_b = N_{\rm nonprompt}/ (N_{\rm nonprompt}+ N_{\rm prompt})$, and they are studied separately for each data-taking year. The first is the $t_z$ resolution model. A sum of three Gaussian functions is used to describe the $t_z$ resolution of the \psitwos and \theX signal.  
As an alternative, a sum of two Gaussian functions is used, and the relative differences of the fitted 
$F_b$ ratio for the \theX and \psitwos signal, \fratio, 
are assigned as systematic uncertainty.
The mean value of the $t_z$ resolution function is fixed to zero in the reference fit.  
However, the reconstructed $t_z$ distribution could be biased, for example due to tracks from \bquark-hadron decays being included in the PV reconstruction. 
The mean value of the resolution function is left to vary freely and the difference of the \fratio ratio is assigned 
as systematic uncertainty.
The long tail of the $t_z$ distribution is due to misassociated primary vertices that can affect the fit result. Instead of using the different-event method, 
the tail is described with a bifurcated exponential with equal slope parameters on the positive and negative sides. 
The relative difference of \fratio values between the two fits 
is used to assign the corresponding uncertainty.
The $t_z$ distribution for the background is obtained using the \sPlot technique. 
The possible correlation between $M_{\jpsipipi}$ and $t_z$ is checked by comparing the signal yields obtained with fits to $M_{\jpsipipi}$ distribution in each $t_z$ bin to those obtained using the \sPlot technique, and is found to be small.
The effects on \fratio is evaluated by fitting the $t_z$ distribution obtained with fits to $M_{\jpsipipi}$ distribution in each $t_z$ bin, and the resulting 2.4\% difference from the nominal one is taken as a systematic uncertainty.
The pseudo-lifetime
of nonprompt \theX candidates is fixed to 1.5\ps in the reference fit, which could affect the fitted $F_b$ fraction. 
As an alternative, the \theX pseudo-lifetime
is fixed to that of the \psitwos meson in each kinematic bin.   
The difference of \fratio ratios between the reference and the alternative fits for each year is assigned as systematic uncertainty.

\begin{table}[tbp]
\caption{Systematic uncertainties of the production cross-section of \theX relative to \psitwos mesons in the kinematic region $4 < \pt < 20$ \gevc and $2.0 < y < 4.5$. Ranges are due to the variation across the $(\pt, y)$ intervals.
When only a range is given for the 13 TeV data it is shared between different data-taking years.
The large ranges in some cases are due to statistical fluctuations of the signal or control samples used to evaluate the uncertainties.
The systematic uncertainties due to the $M_{\jpsipipi}$ and $t_z$ fit  are considered to be $100\,\%$ correlated, and the statistical and systematic uncertainties due to PID and tracking are considered to be uncorrelated.}
\centering
\begin{tabular}{clc|ccc}
\hline
\multicolumn{2}{c}{\multirow{2}{*}{Sources}} &  \multicolumn{4}{c}{Systematic uncertainty (\%)} \\  \cline{3-6}
\multicolumn{2}{c}{}         & 8\tev & \multicolumn{3}{c}{13\tev}  \\
\multicolumn{2}{c}{}         &   2012        & 2016 & 2017 & 2018  \\
\hline
\multirow{6}{*}{Mass fit} &  Signal lineshape & 0.6           & \multicolumn{3}{c}{2.3}  \\
                            & Fraction of two $F_{\rm DSCB}$ & 0.0--3.6           & \multicolumn{3}{c}{0.0--5.6}  \\
                            &  $\sigma_{2}^{\psitwos}/\sigma_{1}^{\psitwos}$ & 0.0--2.7  & \multicolumn{3}{c}{0.0--6.8}  \\
                            & $\sigma_{1}^{\theX}/\sigma_{1}^{\psitwos}$    & 0.2--3.6  & \multicolumn{3}{c}{0.2--5.1}  \\
                            & $\sigma_{2}^{\theX}/\sigma_{2}^{\psitwos}$    & 0.2--5.6  & \multicolumn{3}{c}{0.2--6.2}  \\
                            & Background lineshape & 0.0--1.5  & \multicolumn{3}{c}{0.0--3.7}  \\
\hline
\multirow{5}{*}{$t_z$ fit} & $t_z$ resolution function    & 0.0--1.4  & 0.0--3.0  & 0.0--1.6 & 0.0--1.0 \\
                              &Fixed mean of $t_z$ resolution & 0.0--0.4  & 0.0--1.0 & 0.0--0.8 & 0.0--0.6\\
                              &Wrong PV             & 0.0--2.8  & 0.0--4.1  & 0.0--3.6 & 0.0--1.8 \\
                              & Backgrond shape & 2.4  & \multicolumn{3}{c}{2.4} \\
                              &Fixed pseudo-lifetime  & 0.1--10.9  & 1.0--12.1 & 1.3--8.0 & 1.3--7.8\\
\hline
 \multicolumn{2}{l}{Tracking}  & 0.1--0.7 &  0.1--0.2  & 0.1--1.4 & 0.1--1.0  \\
 \multicolumn{2}{l}{Muon identification}   &  0.0--6.1 & 0.0--1.8  & 0.0--1.8 & 0.0--1.5 \\
 \multicolumn{2}{l}{Pion identification}   &  0.1--6.7 & 0.0--0.4  & 0.0--0.9 & 0.0--0.3 \\
 \multicolumn{2}{l}{Trigger thresholds}   &  -- & 0.0--15.1  & 0.3--6.4 & 0.3--7.3 \\
 \multicolumn{2}{l}{Simulation weighting}   &  4.5--9.3 & 3.6--7.4  & 3.2--8.9 & 3.2--6.1 \\
 \multicolumn{2}{l}{Global event requirements}   &  0.5 & \multicolumn{3}{c}{1.9} \\
 \multicolumn{2}{l}{$M_{\pi^{+}\pi^{-}}$ spectrum}   &  2.0 & \multicolumn{3}{c}{2.0}  \\
 \multicolumn{2}{l}{Trigger efficiency}   &  1.0& \multicolumn{3}{c}{1.0}  \\
\hline
\multicolumn{2}{l}{Total systematic uncertainty}  &  6.7--14.8 & 7.1--17.9   & 6.0--15.3 & 6.0--13.1  \\
\hline
\multicolumn{2}{l}{Total statistical uncertainty: prompt}   &  7--17         & 5--19     &  6--31  & 5--13 \\
\multicolumn{2}{l}{Total statistical uncertainty: nonprompt}   &  13--26 & 11--23    &  10--32 & 9--19 \\
\hline
\end{tabular}\label{tab:sys_sum}
\end{table}

The track detection efficiencies are determined from a simulated sample in each $(\pt,y)$ interval, 
and are corrected using control data.  
The statistical uncertainty due to the limited size of the control data sample is propagated using a large number of pseudoexperiments. For each pseudoexperiment, a new efficiency-correction ratio as a function of the $(\pt,y)$ interval is generated according to a Gaussian distribution where the original efficiency ratio and its uncertainty are used as the Gaussian mean and standard deviation, respectively. 

The systematic uncertainty due to particle identification is studied considering the following contributions. 
The first is the statistical uncertainty due to the limited size of the calibration sample, 
which is estimated using pseudoexperiments and 
found to be negligible compared to other systematic uncertainties. 
The second is due to the binning scheme of the calibration sample.
This contribution is studied by varying the binning in momentum, pseudorapidity and track multiplicity. 
The maximum differences among these contributions on the efficiency ratios are taken as systematic uncertainty.

The hardware-trigger thresholds on muon and hadron \pt varied throughout data taking, however only one value is used in the simulation.
Differences in the trigger efficiencies observed when varying the thresholds in the simulation are taken as a source of systematic uncertainty.  
The $\pt$ and $y$ distributions of the simulated $\theX$ and $\psitwos$ samples are corrected to match those in the data.
The uncertainty on the simulation weighting is studied by propagating the statistical uncertainty on the correction using 
the bootstrap method~\cite{efron1982jackknife}. The bootstrap method is used to generate 100 pseudoexperiments according to the data sample. The simulation weightings are performed with the generated pseudoexperiments. The efficiencies are calculated for each weighting, and the root-mean-square of the resulting efficiency distribution is taken as systematic uncertainty. 
The effects of the global event requirements are estimated through the difference of the \effratio ratio between the data and the simulation. 
The $M_{\pi^{+}\pi^{-}}$ distributions in the data and the simulation are slightly different, especially in the high $M_{\pi^{+}\pi^{-}}$ region. 
This difference affects the \effratio ratio and is taken as systematic uncertainty. 
The systematic uncertainty on the trigger efficiency is taken from \jpsi pair production measurement~\cite{LHCb-PAPER-2016-057}.

\section{Results}
\label{sec:results}

\begin{figure}
  \begin{center}
    \includegraphics[width=0.49\linewidth]{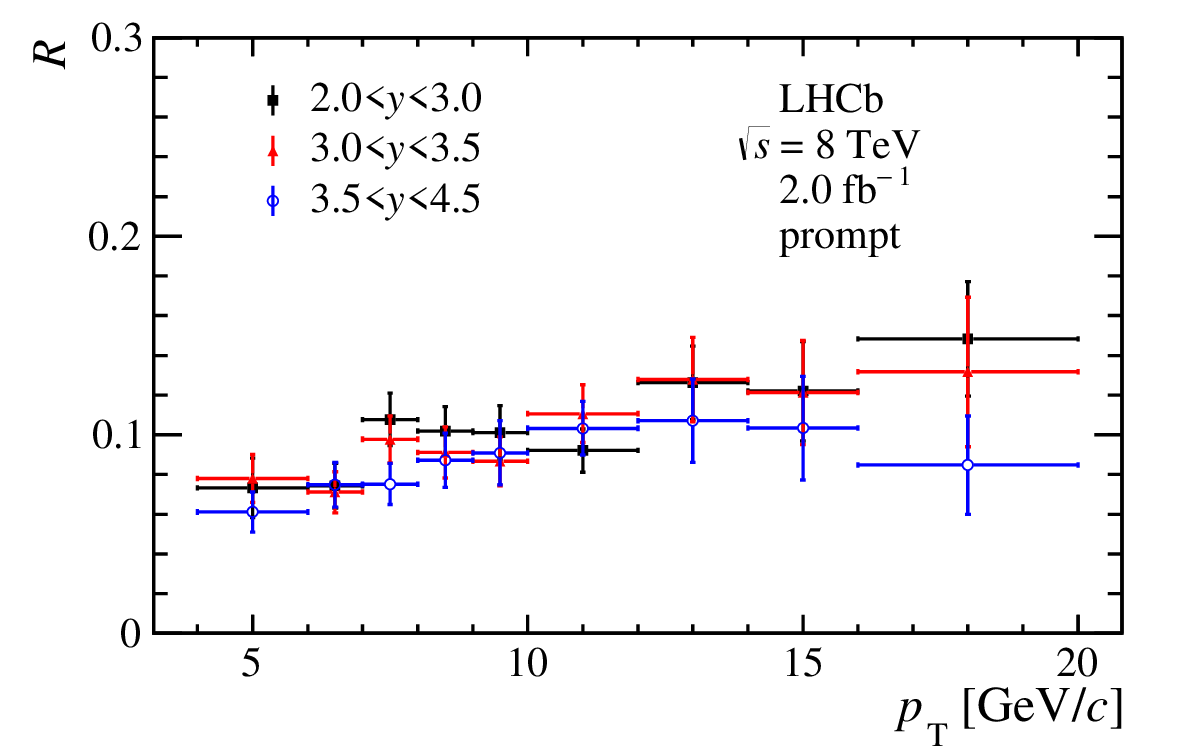} 
    \includegraphics[width=0.49\linewidth]{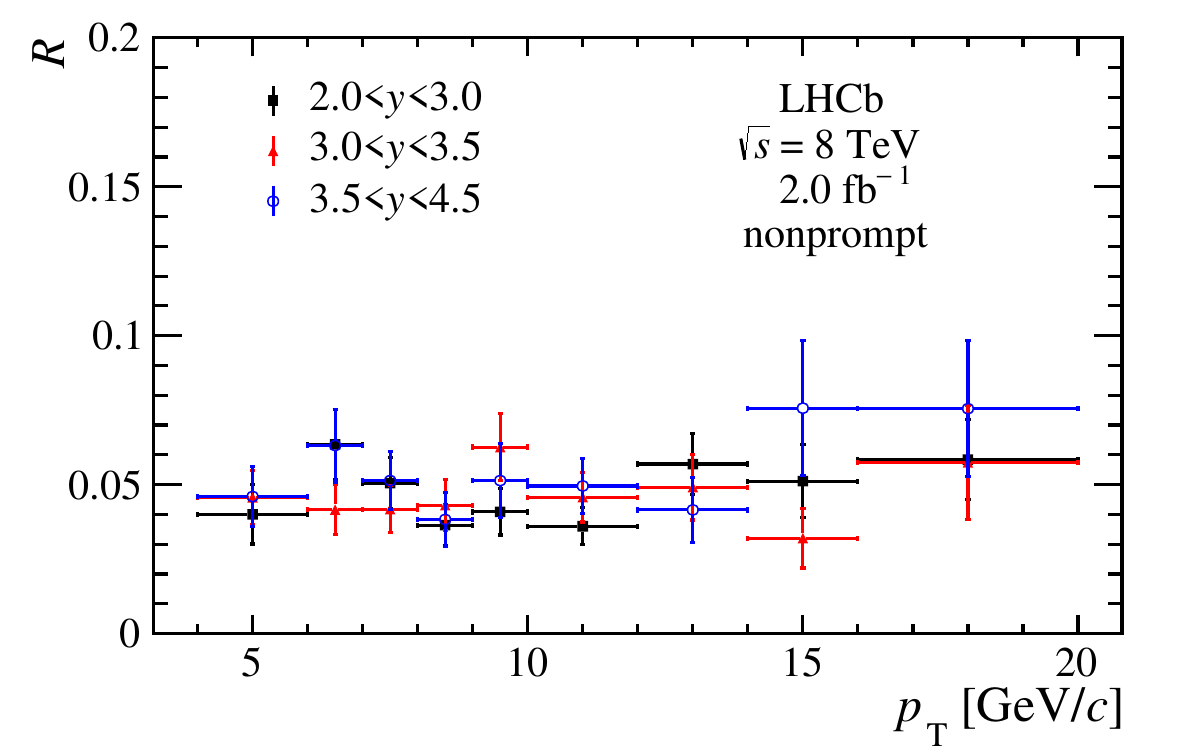} 
    \vspace*{-0.5cm}
  \end{center}
  \caption{Ratios of differential cross-section times the branching ratio to the \jpsipipi final state between \theX and \psitwos mesons from (left) prompt production and (right) nonprompt from \bquark-hadron decays, for the \sqrts{8} sample as a function of \pt in intervals of rapidity.
In all panels, the error bars represent the sum in quadrature of the statistical and systematic uncertainties.
    }
  \label{fig:cs_ratio_12}
\end{figure}

\begin{figure}
  \begin{center}
    \includegraphics[width=0.49\linewidth]{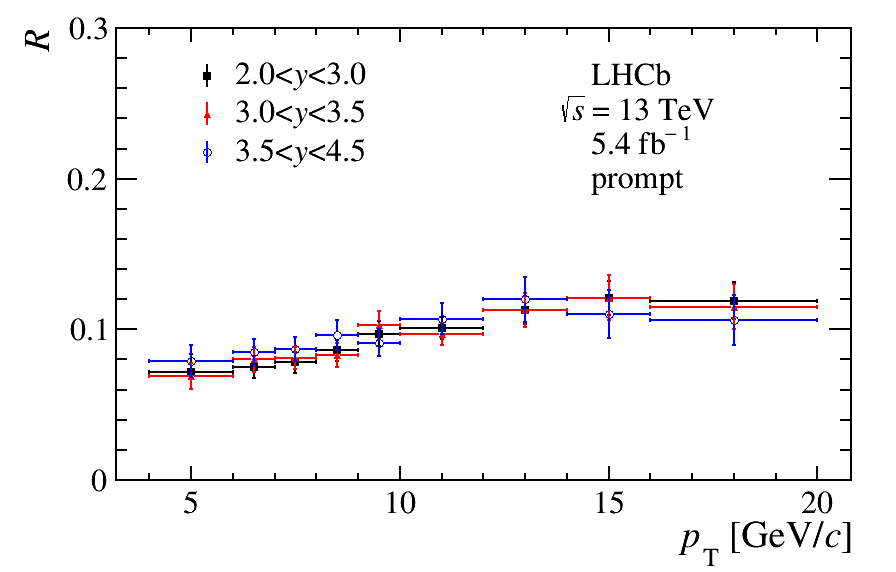} 
    \includegraphics[width=0.49\linewidth]{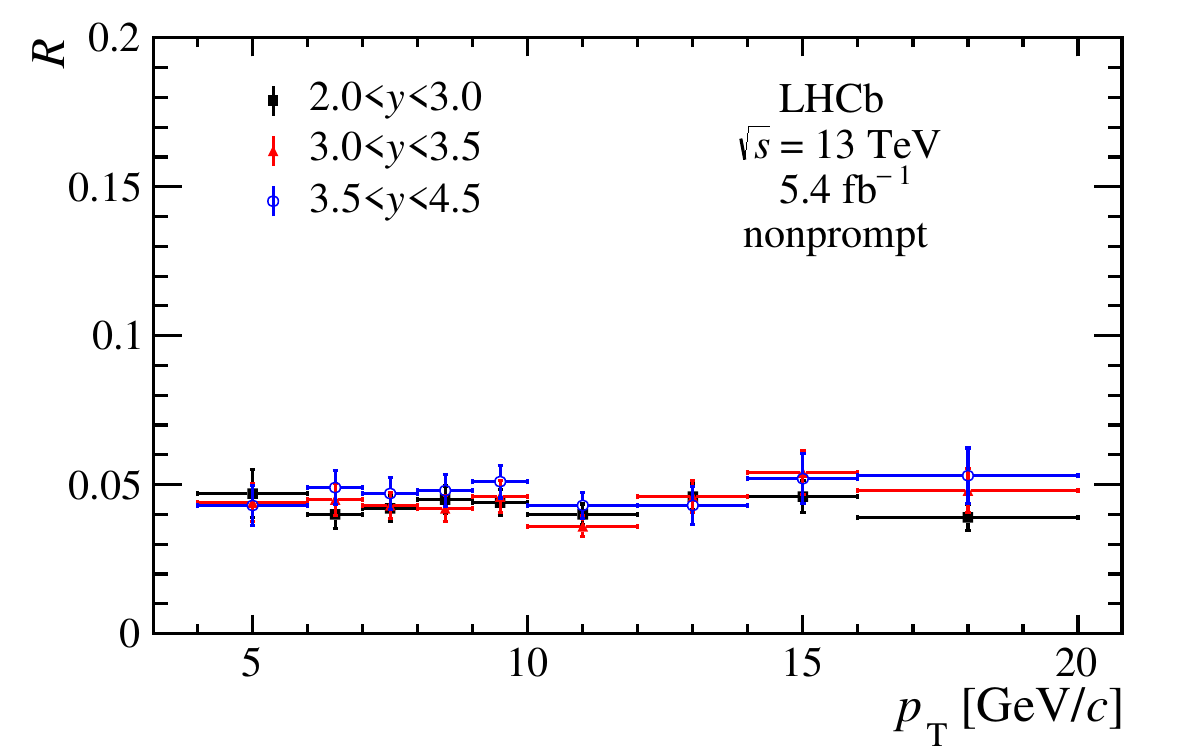} 
    \vspace*{-0.5cm}
  \end{center}
  \caption{Ratios of differential cross-section times the branching ratio to the \jpsipipi final state between \theX and \psitwos mesons from (left) prompt production and (right) nonprompt from \bquark-hadron decays, for the \sqrts{13} sample as a function of \pt in intervals of rapidity. In all panels, the error bars represent the sum in quadrature of the statistical and systematic uncertainties.
    }
  \label{fig:cs_ratio_run2}
\end{figure}

The double-differential cross-section of the \theX state relative to that of the \psitwos meson is measured as a function of \pt and $y$ using $pp$ collision data taken at centre-of-mass energies of $\sqrt{s}=8$ and 13\tev. 
The analysis assumes unpolarised production (a study on the impact of polarisation is described in Appendix~\ref{sec:polar}).
For the per-year measurements at 13\tev, the combination of the cross-section ratios is performed using the Best Linear Unbiased Estimate (BLUE) method~\cite{Lyons:1988rp,Valassi:2003mu,Valassi:2013bga}.    
The weighted average of these measurements is calculated by minimising the total uncertainty of the result and accounting for correlations between per-year measurements. 
The cross-section ratios for promptly and nonpromptly produced mesons
measured with the 8 and 13\tev data samples as a function of \pt and $y$ are shown in Figs.~\ref{fig:cs_ratio_12} and~\ref{fig:cs_ratio_run2}, respectively.
The cross-section times branching ratios of the \theX over \psitwos, integrated over the kinematic region $4 < \pt < 20$ \gevc and $2.0 < y < 4.5$, are obtained to be 
\begin{align*}
 R_{\rm{prompt}}^{\rm{8\,TeV}} &= (7.6\pm0.5\pm0.9) \times 10^{-2},\\
 R_{\rm{nonprompt}}^{\rm{8\,TeV}} &= (4.6\pm0.4\pm0.5) \times 10^{-2},\\
 R_{\rm{prompt}}^{\rm{13\,TeV}} &= (7.6\pm0.3\pm0.6) \times 10^{-2},\\
 R_{\rm{nonprompt}}^{\rm{13\,TeV}} &= (4.4\pm0.2\pm0.4) \times 10^{-2},
\end{align*}
where the first uncertainties are statistical and the second systematic.

\begin{figure}
  \begin{center}
    \includegraphics[width=0.5\linewidth]{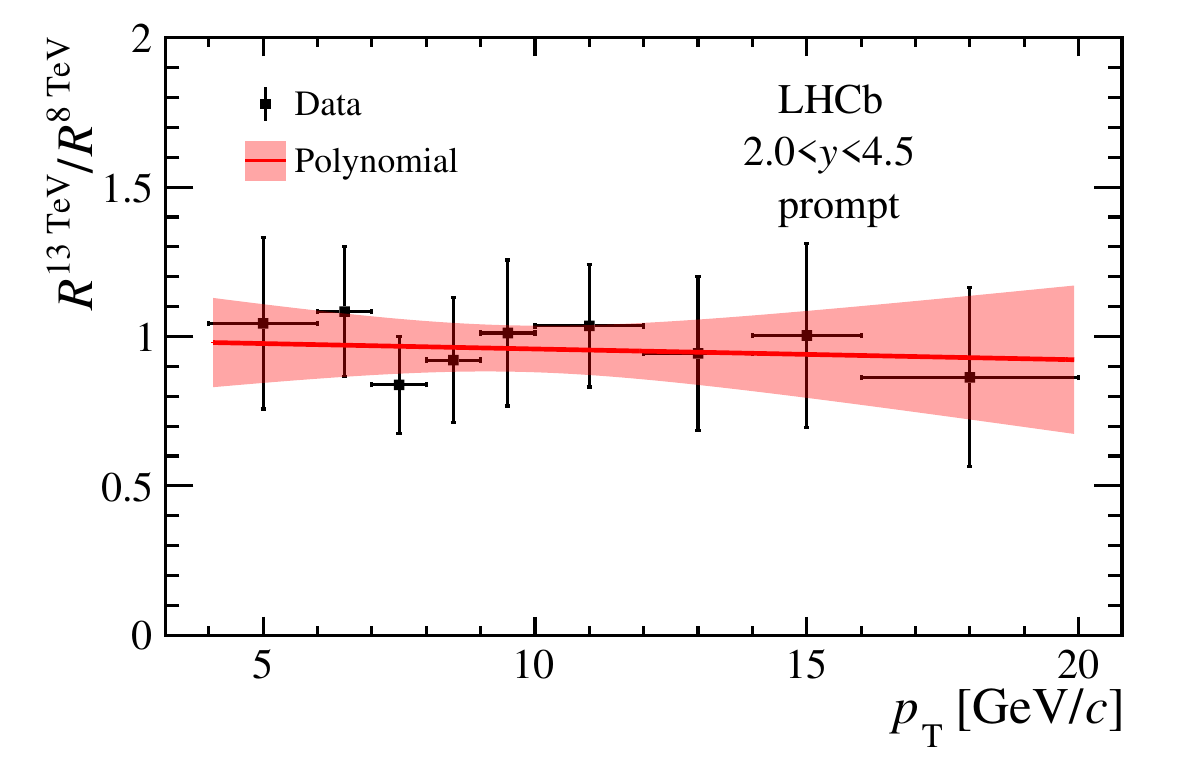} 
    \vspace*{-0.5cm}
  \end{center}
  \caption{
Double ratio of the prompt \theX production cross-section relative to that of \psitwos mesons between 13 and 8\tev as a function of \pt integrated over $2.0<y<4.5$. 
The red line with the solid band represent the fit result to a first-order polynomial and its uncertainty. 
    }
  \label{fig:cs_double_ratio}
\end{figure}

The double ratio of the prompt \theX and \psitwos production cross-sections between 13 and 8\tev is also calculated using the measured cross-section ratio for 8\tev 
and the combined ratio for 13\tev.  
Figure~\ref{fig:cs_double_ratio} shows the double ratio of production cross-sections as a function of \pt integrated over the kinematic region $2.0<y<4.5$. 
A first-order polynomial of the form $R^{\rm{13\,TeV}}/R^{\rm{8\,TeV}}=a_0+a_1\pt$ 
is used to fit the double ratio, yielding $a_{0} = 0.99 \pm 0.23$
and a slope of $a_{1} = (4\pm23)\times10^{-3} (\gevc)^{-1}$,  consistent with zero.

\section{Conclusion}
\label{sec:conclusion}
In summary, the production cross-section of the \theX state relative to the \psitwos meson is measured using $pp$ data collected at centre-of-mass energies of 8 and 13\tev.
The double-differential cross-section ratio times their ratio of branching fractions to the \jpsipipi final state, as a function of \pt and $y$ in the ranges $4<\pt<20$ \gevc and $2.0<y<4.5$, are determined for prompt and nonprompt production of \theX states relative to \psitwos mesons.
The prompt ratio increases as a function of \pt, showing that the \theX  production is less suppressed relative to the one of prompt \psitwos mesons in the higher \pt region. 
This behaviour is similar to the case of prompt production of \psitwos relative to \jpsi mesons as measured by the CMS~\cite{CMS:2011kxe} and LHCb experiments~\cite{LHCB-PAPER-2011-045}, and is consistent with theoretical predictions~\cite{Khoze:2004eu}. 
Using the production cross-section of the \psitwos meson measured by the LHCb experiment at 13 \tev~\cite{LHCb-PAPER-2018-049}, the absolute
production cross-section of the \theX meson at 13 \tev multiplied by its branching fraction to the \jpsipipi final state is determined as a  function of \pt, as detailed in Appendix~\ref{sec:abs_cs}. The result is found to agree in the $\pt>10\gevc$ region with NLO NRQCD predictions~\cite{Meng:2013gga}, which model the \theX state as a mixture of $\chi_{c1}(2P)$ and $D^0\Dstarzb$ molecule states, produced through its $\chi_{c1}(2P)$ component.
The prompt cross-section ratios at 13 and 8\tev are also compared, and no significant dependence on the centre-of-mass energy is found. 
The nonprompt ratios of cross-sections at 13 and 8\tev are consistent with a flat distribution, 
determined by the \bquark-decay branching ratios.

\section*{Acknowledgements}
\noindent We express our gratitude to our colleagues in the CERN
accelerator departments for the excellent performance of the LHC. We
thank the technical and administrative staff at the LHCb
institutes.
We acknowledge support from CERN and from the national agencies:
CAPES, CNPq, FAPERJ and FINEP (Brazil); 
MOST and NSFC (China); 
CNRS/IN2P3 (France); 
BMBF, DFG and MPG (Germany); 
INFN (Italy); 
NWO (Netherlands); 
MNiSW and NCN (Poland); 
MEN/IFA (Romania); 
MSHE (Russia); 
MICINN (Spain); 
SNSF and SER (Switzerland); 
NASU (Ukraine); 
STFC (United Kingdom); 
DOE NP and NSF (USA).
We acknowledge the computing resources that are provided by CERN, IN2P3
(France), KIT and DESY (Germany), INFN (Italy), SURF (Netherlands),
PIC (Spain), GridPP (United Kingdom), RRCKI and Yandex
LLC (Russia), CSCS (Switzerland), IFIN-HH (Romania), CBPF (Brazil),
PL-GRID (Poland) and NERSC (USA).
We are indebted to the communities behind the multiple open-source
software packages on which we depend.
Individual groups or members have received support from
ARC and ARDC (Australia);
AvH Foundation (Germany);
EPLANET, Marie Sk\l{}odowska-Curie Actions and ERC (European Union);
A*MIDEX, ANR, IPhU and Labex P2IO, and R\'{e}gion Auvergne-Rh\^{o}ne-Alpes (France);
Key Research Program of Frontier Sciences of CAS, CAS PIFI, CAS CCEPP, 
Fundamental Research Funds for the Central Universities, 
and Sci. \& Tech. Program of Guangzhou (China);
RFBR, RSF and Yandex LLC (Russia);
GVA, XuntaGal and GENCAT (Spain);
the Leverhulme Trust, the Royal Society
 and UKRI (United Kingdom).

\clearpage
% ===============================================================================
% Purpose: appendix to the standard template: standard symbol alises from Ulrik
% Author: Tomasz Skwarnicki
% Created on: 2009-09-24
% ===============================================================================

%{\noindent\normalfont\bfseries\Large Appendices}
\section*{Appendices}

\appendix
\section{\boldmath Polarisation of \texorpdfstring{\psitwos}{psi(2S)} and \texorpdfstring{\theX}{chi-c1(3872} mesons}
\label{sec:polar}

The polarisation of the \jpsi meson is directly inherited from the \psitwos parent, since the dipion system is produced in an $S$-wave state
relative to the $\jpsi$ meson.
For the \mbox{$\theX\to\jpsipipi$} decay, the determination of the \theX polarisation can be obtained by
measuring the dimuon angular decay distribution in the rest frame of the \jpsi daughter, as discussed in  Refs.~\cite{Sirunyan:2019apc,Faccioli:2011be}.
The angular dependence of the $\jpsi\rightarrow\mu^{+}\mu^{-}$ decay for \theX and \psitwos mesons is
\begin{eqnarray}
 \label{eq:angle_psi}
\frac{d^{2}N}{d\cos\theta d\phi} \propto 1 + \lambda_{\theta}\cos^{2}\theta + \lambda_{\phi}\sin^{2}\theta
\cos2\phi + \lambda_{\theta\phi}\sin2\theta\cos\phi \,,
\end{eqnarray}
where $\lambda_{i}$ are the polarisation parameters and $\theta (\phi)$ are the polar (azimuthal) angles between the positively charged muon in the
$\jpsi\rightarrow\mu^{+}\mu^{-}$ rest frame and the direction of the \psitwos meson in the laboratory frame. Various polarisation hypotheses are considered:
\begin{itemize}
\item Unpolarised, with an isotropic distribution that is independent of the polarisation parameters, $\lambda_{\theta} = \lambda_{\phi} = \lambda_{\theta\phi} = 0$. This is used as the central hypothesis.

\item Transversely polarised with $\lambda_{\theta} = +1, \lambda_{\phi} = \lambda_{\theta\phi} = 0$, labelled as $T_{+0}$.

\item Transversely polarised with $\lambda_{\theta} = +1, \lambda_{\phi} = +1, \lambda_{\theta\phi} = 0$, labelled as $T_{++}$.

\item Transversely polarised with $\lambda_{\theta} = +1, \lambda_{\phi} = -1, \lambda_{\theta\phi} = 0$, labelled as $T_{+-}$.

\item Longitudinaly polarised, with the parameters $\lambda_{\theta} = -1, \lambda_{\phi} = \lambda_{\theta\phi} = 0$, labelled as $L$.

\item Off-Plane Positive, with the polarisation parameters $\lambda_{\theta} = 0, \lambda_{\phi} = 0, \lambda_{\theta\phi} = +0.5$, labelled as $OP+$.

\item Off-Plane Negative, with the polarisation parameters $\lambda_{\theta} = 0, \lambda_{\phi} = 0, \lambda_{\theta\phi} = -0.5$, labelled as $OP-$.
\end{itemize}

\begin{figure}[!htbp]
  \begin{center}
    \includegraphics[width=0.49\linewidth]{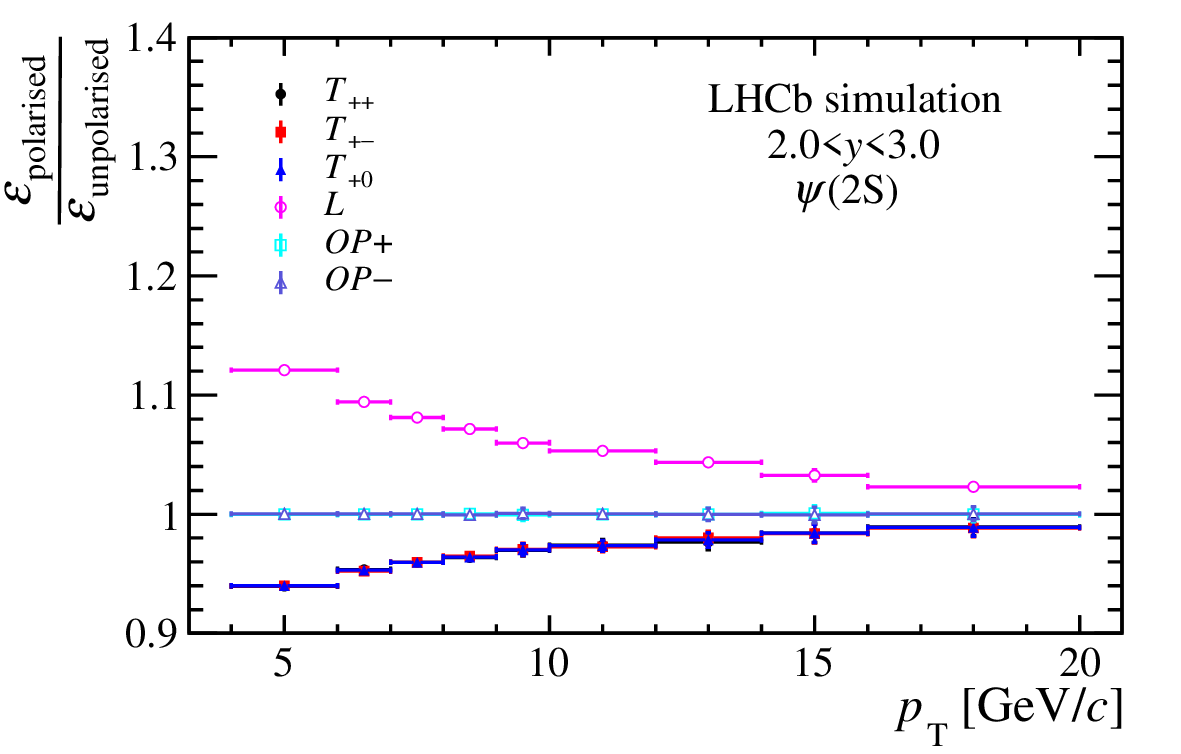}
    \includegraphics[width=0.49\linewidth]{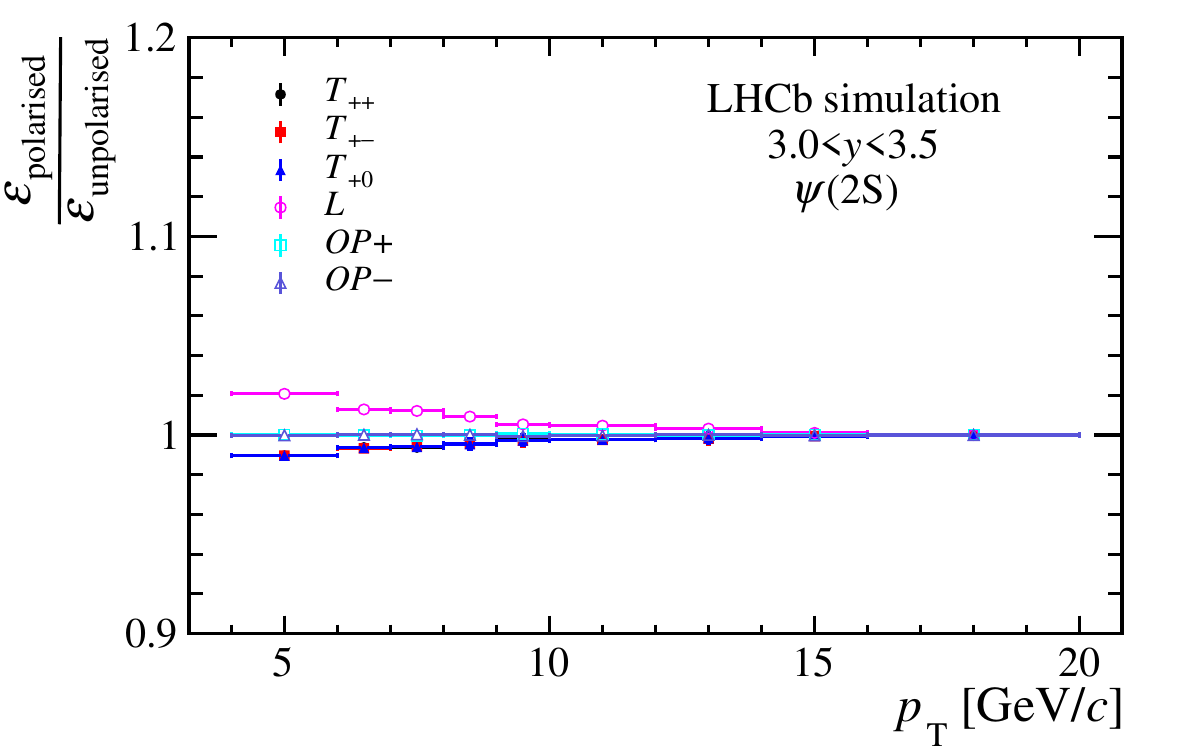}
    \\
    \includegraphics[width=0.49\linewidth]{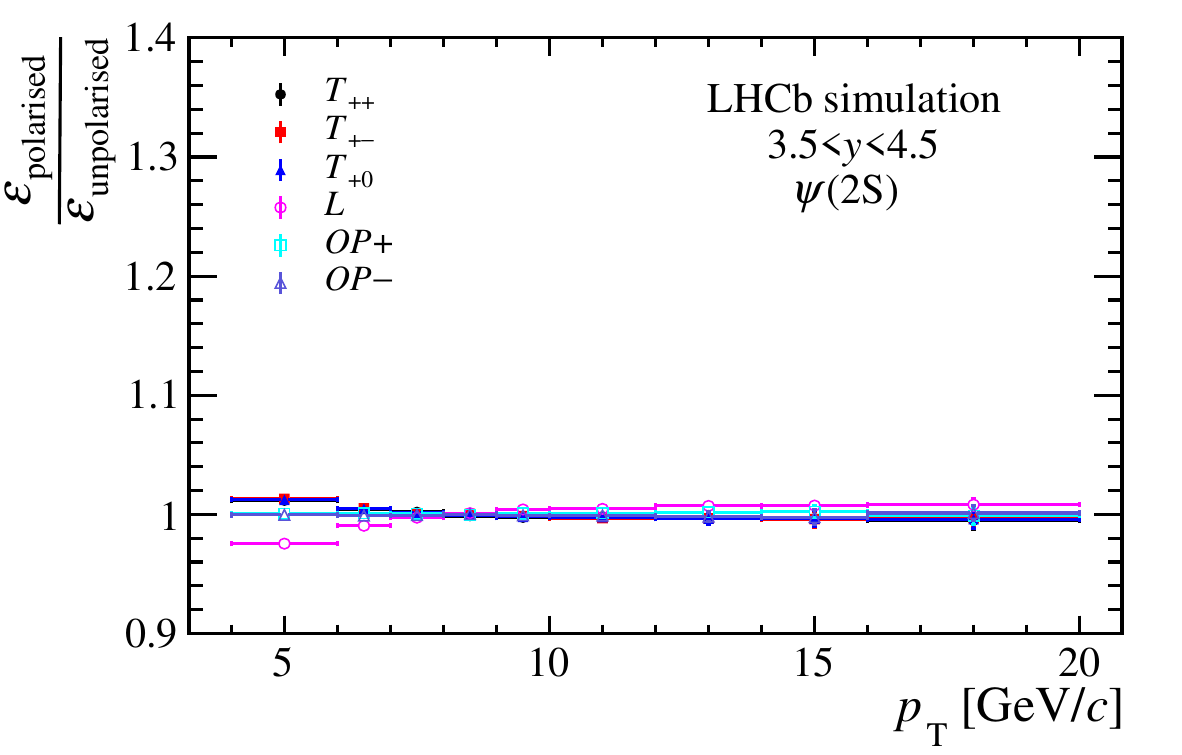}
    \vspace*{-0.5cm}
  \end{center}
  \caption{
Ratio of acceptance efficiencies for \psitwos mesons for various polarisation hypotheses with respect to the unpolarised case.
}
  \label{fig:ratio_acc_psi}
\end{figure}

\begin{figure}[!htbp]
  \begin{center}
    \includegraphics[width=0.49\linewidth]{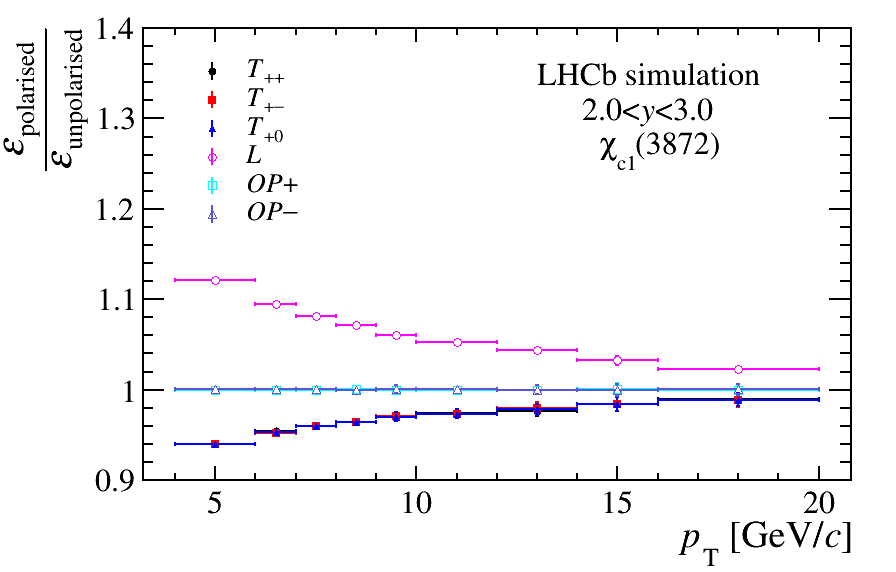}
    \includegraphics[width=0.49\linewidth]{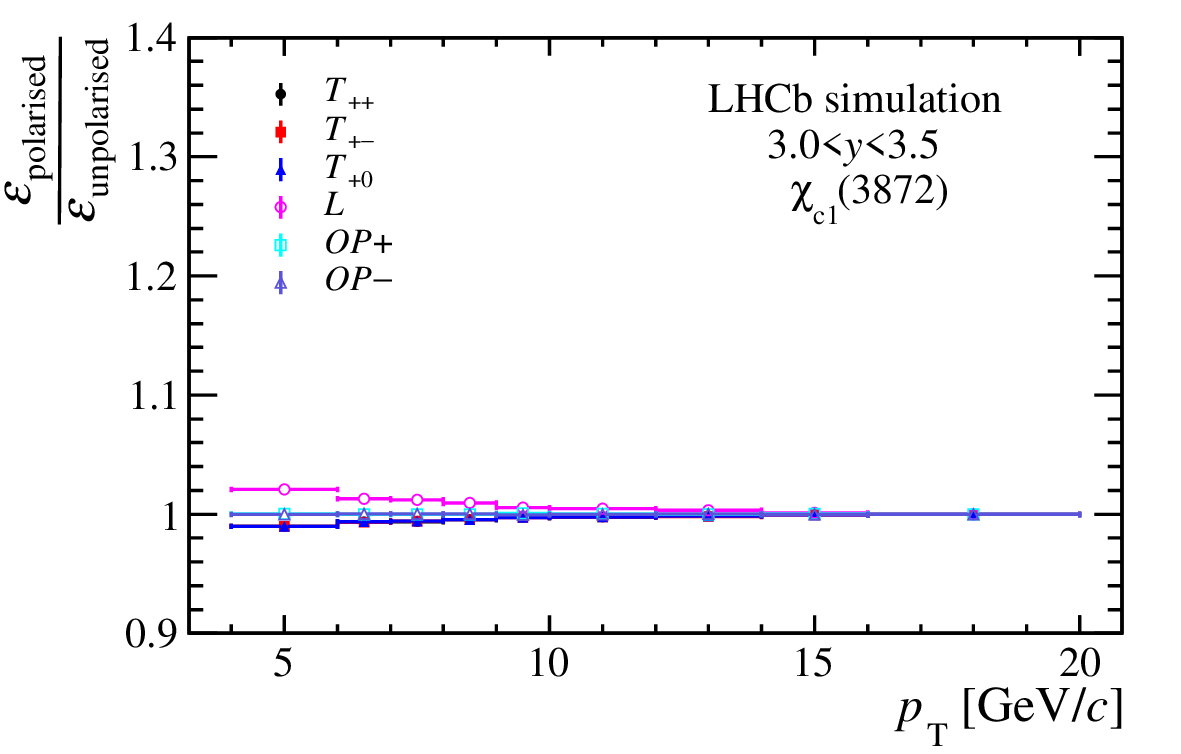}
\\
    \includegraphics[width=0.49\linewidth]{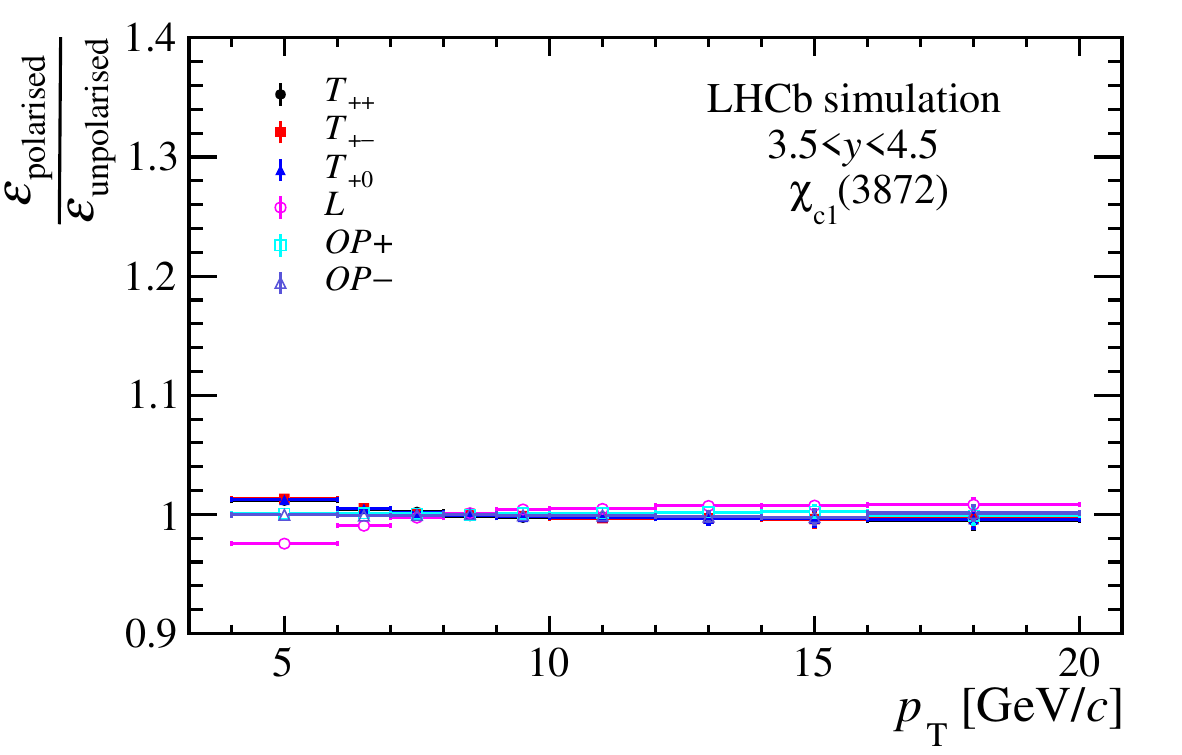}
    \vspace*{-0.5cm}
  \end{center}
  \caption{
Ratio of acceptance efficiencies for \theX mesons for various polarisation hypotheses with respect to the unpolarised case.
}
  \label{fig:ratio_acc_x}
\end{figure}

The acceptance weights are calculated for each of these scenarios in each $(\pt,y)$ interval.
The ratios of the acceptance efficiencies for each polarisation scenario to those of the unpolarised case
are shown in Fig.~\ref{fig:ratio_acc_psi} for \psitwos mesons and  Fig.~\ref{fig:ratio_acc_x} for the \theX state, and the values are listed in
Tables~\ref{tab:tab_ratio_acc_psi_y1},~\ref{tab:tab_ratio_acc_psi_y2}, and~\ref{tab:tab_ratio_acc_psi_y3} for the former, and
Tables~\ref{tab:tab_ratio_acc_x_y1},~\ref{tab:tab_ratio_acc_x_y2}, and~\ref{tab:tab_ratio_acc_x_y3} for the latter.

\begin{table}[!htbp]
\begin{center}
\caption{
Ratio of acceptance efficiencies for \psitwos mesons for various polarisation hypotheses with respect to the unpolarised case, in the interval $2.0<y<3.0$.
}
\label{tab:tab_ratio_acc_psi_y1}
\begin{tabular}{c|cccccc}
\hline
$$\pt[\gevc]$$ &  $T_{+0}$ & $T_{++}$ & $T_{+-}$ & $L$ & $OP+$ & $OP-$  \\ \hline
4-6 & 0.940 & 0.940 & 0.940 & 1.121 & 1.000 & 1.000\\
6-7 & 0.953 & 0.954 & 0.952 & 1.094 & 1.000 & 1.000\\
7-8 & 0.960 & 0.959 & 0.960 & 1.081 & 1.000 & 1.000\\
8-9 & 0.964 & 0.964 & 0.965 & 1.072 & 1.000 & 1.000\\
9-10 & 0.970 & 0.970 & 0.970 & 1.060 & 0.999 & 1.001\\
10-12 & 0.974 & 0.974 & 0.973 & 1.053 & 1.000 & 1.000\\
12-14 & 0.978 & 0.977 & 0.980 & 1.043 & 1.000 & 1.000\\
14-16 & 0.984 & 0.984 & 0.984 & 1.033 & 1.001 & 0.999\\
16-20 & 0.989 & 0.989 & 0.988 & 1.023 & 1.000 & 1.000\\
\hline
\end{tabular}
\end{center}
\end{table}

\begin{table}[!htbp]
\begin{center}
\caption{
Ratio of acceptance efficiencies for \psitwos mesons for various polarisation hypotheses with respect to the unpolarised case, in the interval $3.0<y<3.5$.
}
\label{tab:tab_ratio_acc_psi_y2}
\begin{tabular}{c|cccccc}
\hline
$$\pt[\gevc]$$ &  $T_{+0}$ & $T_{++}$ & $T_{+-}$ & $L$ & $OP+$ & $OP-$  \\ \hline
4-6 & 0.990 & 0.990 & 0.990 & 1.021 & 1.000 & 1.000\\
6-7 & 0.994 & 0.994 & 0.993 & 1.013 & 1.000 & 1.000\\
7-8 & 0.994 & 0.994 & 0.994 & 1.012 & 1.000 & 1.000\\
8-9 & 0.995 & 0.995 & 0.996 & 1.009 & 1.000 & 1.000\\
9-10 & 0.997 & 0.997 & 0.997 & 1.005 & 1.001 & 0.999\\
10-12 & 0.998 & 0.998 & 0.998 & 1.005 & 1.000 & 1.000\\
12-14 & 0.998 & 0.999 & 0.998 & 1.003 & 1.000 & 1.000\\
14-16 & 1.000 & 1.000 & 0.999 & 1.001 & 1.000 & 1.000\\
16-20 & 1.000 & 1.000 & 1.000 & 1.000 & 1.000 & 1.000\\
\hline
\end{tabular}
\end{center}
\end{table}

\begin{table}[!htbp]
\begin{center}
\caption{
Ratio of acceptance efficiencies for \psitwos mesons for various polarisation hypotheses with respect to the unpolarised case, in the interval $3.5<y<4.5$.
}
\label{tab:tab_ratio_acc_psi_y3}
\begin{tabular}{c|cccccc}
\hline
$$\pt[\gevc]$$ &  $T_{+0}$ & $T_{++}$ & $T_{+-}$ & $L$ & $OP+$ & $OP-$  \\ \hline
4-6 & 1.012 & 1.012 & 1.013 & 0.975 & 1.000 & 1.000\\
6-7 & 1.005 & 1.005 & 1.005 & 0.990 & 1.001 & 0.999\\
7-8 & 1.001 & 1.002 & 1.001 & 0.997 & 1.001 & 0.999\\
8-9 & 0.999 & 0.999 & 1.000 & 1.001 & 1.000 & 1.000\\
9-10 & 0.998 & 0.997 & 0.999 & 1.004 & 1.001 & 0.999\\
10-12 & 0.998 & 0.998 & 0.997 & 1.005 & 1.001 & 0.999\\
12-14 & 0.996 & 0.996 & 0.997 & 1.007 & 1.002 & 0.998\\
14-16 & 0.996 & 0.997 & 0.996 & 1.007 & 1.002 & 0.998\\
16-20 & 0.996 & 0.995 & 0.997 & 1.008 & 0.999 & 1.001\\
\hline
\end{tabular}
\end{center}
\end{table}

\begin{table}[!htbp]
\begin{center}
\caption{
Ratio of acceptance efficiencies for \theX mesons for various polarisation hypotheses with respect to the unpolarised case, in the interval $2.0<y<3.0$.
}
\label{tab:tab_ratio_acc_x_y1}
\begin{tabular}{c|cccccc}
\hline
$$\pt[\gevc]$$ &  $T_{+0}$ & $T_{++}$ & $T_{+-}$ & $L$ & $OP+$ & $OP-$  \\ \hline
4-6 & 0.941 & 0.941 & 0.941 & 1.118 & 1.000 & 1.000\\
6-7 & 0.952 & 0.952 & 0.952 & 1.096 & 1.000 & 1.000\\
7-8 & 0.958 & 0.958 & 0.958 & 1.084 & 1.000 & 1.000\\
8-9 & 0.964 & 0.963 & 0.965 & 1.071 & 1.000 & 1.000\\
9-10 & 0.968 & 0.968 & 0.968 & 1.063 & 0.999 & 1.001\\
10-12 & 0.973 & 0.972 & 0.973 & 1.055 & 1.000 & 1.000\\
12-14 & 0.977 & 0.978 & 0.975 & 1.047 & 1.000 & 1.000\\
14-16 & 0.985 & 0.984 & 0.986 & 1.030 & 1.000 & 1.000\\
16-20 & 0.986 & 0.987 & 0.986 & 1.027 & 1.001 & 0.999\\
\hline
\end{tabular}
\end{center}
\end{table}

\begin{table}[!htbp]
\begin{center}
\caption{
Ratio of acceptance efficiencies for \theX mesons for various polarisation hypotheses with respect to the unpolarised case, in the interval $3.0<y<3.5$.
}
\label{tab:tab_ratio_acc_x_y2}
\begin{tabular}{c|cccccc}
\hline
$$\pt[\gevc]$$ &  $T_{+0}$ & $T_{++}$ & $T_{+-}$ & $L$ & $OP+$ & $OP-$  \\ \hline
4-6 & 0.990 & 0.990 & 0.990 & 1.021 & 1.000 & 1.000\\
6-7 & 0.992 & 0.992 & 0.993 & 1.015 & 1.000 & 1.000\\
7-8 & 0.994 & 0.994 & 0.995 & 1.012 & 1.000 & 1.000\\
8-9 & 0.995 & 0.995 & 0.996 & 1.009 & 1.000 & 1.000\\
9-10 & 0.997 & 0.998 & 0.997 & 1.005 & 0.999 & 1.001\\
10-12 & 0.997 & 0.997 & 0.998 & 1.005 & 1.000 & 1.000\\
12-14 & 0.999 & 0.999 & 0.999 & 1.002 & 1.000 & 1.000\\
14-16 & 0.998 & 0.998 & 0.998 & 1.003 & 1.000 & 1.000\\
16-20 & 0.999 & 0.999 & 0.999 & 1.002 & 1.001 & 0.999\\
\hline
\end{tabular}
\end{center}
\end{table}

\begin{table}[!htbp]
\begin{center}
\caption{
Ratio of acceptance efficiencies for \theX mesons for various polarisation hypotheses with respect to the unpolarised case, in the interval $3.5<y<4.5$.
}
\label{tab:tab_ratio_acc_x_y3}
\begin{tabular}{c|cccccc}
\hline
$$\pt[\gevc]$$ &  $T_{+0}$ & $T_{++}$ & $T_{+-}$ & $L$ & $OP+$ & $OP-$  \\ \hline
4-6 & 1.014 & 1.014 & 1.013 & 0.972 & 1.000 & 1.000\\
6-7 & 1.006 & 1.006 & 1.006 & 0.988 & 1.001 & 0.999\\
7-8 & 1.003 & 1.003 & 1.003 & 0.995 & 1.001 & 0.999\\
8-9 & 0.999 & 0.998 & 1.000 & 1.002 & 1.000 & 1.000\\
9-10 & 0.999 & 0.999 & 0.999 & 1.002 & 1.001 & 0.999\\
10-12 & 0.996 & 0.997 & 0.996 & 1.007 & 1.001 & 0.999\\
12-14 & 0.996 & 0.996 & 0.995 & 1.008 & 1.001 & 0.999\\
14-16 & 0.997 & 0.996 & 0.997 & 1.006 & 0.999 & 1.001\\
16-20 & 0.998 & 0.998 & 0.998 & 1.004 & 1.000 & 1.000\\
\hline
\end{tabular}
\end{center}
\end{table}

\clearpage

\section{\boldmath Absolute cross-section of \theX}
\label{sec:abs_cs}
\begin{figure}[!hb]
  \begin{center}
    \includegraphics[width=0.49\linewidth]{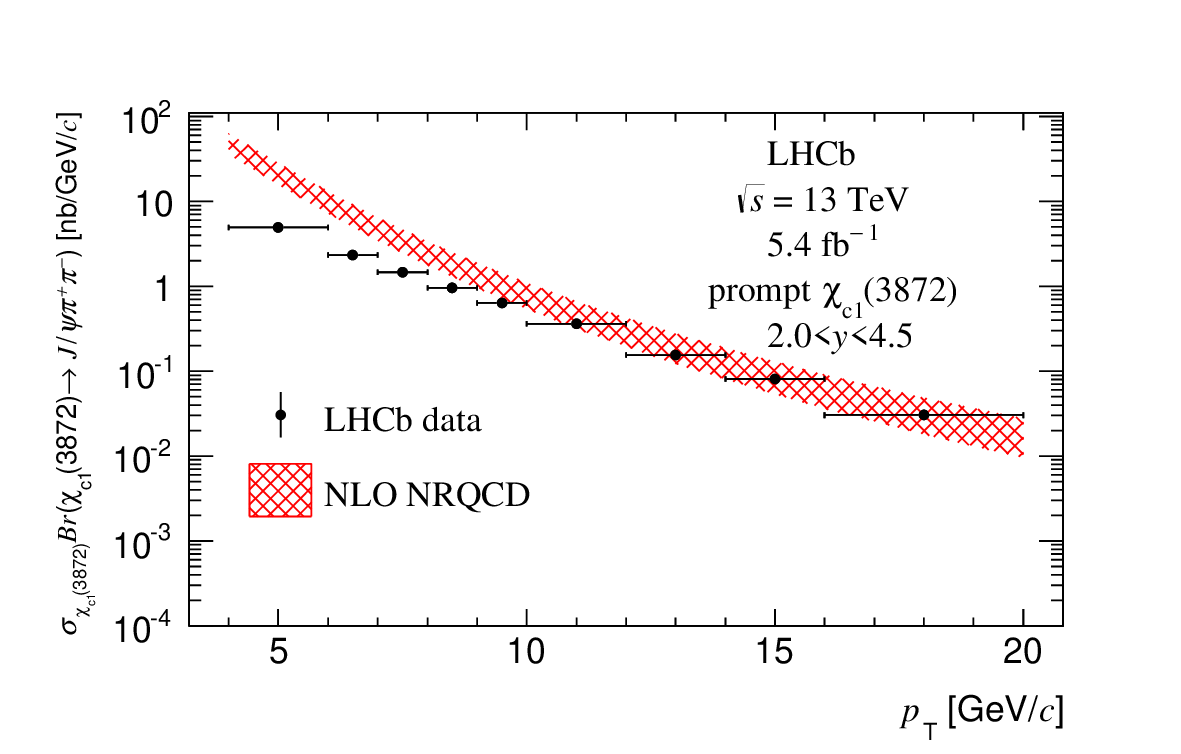} 
    \includegraphics[width=0.49\linewidth]{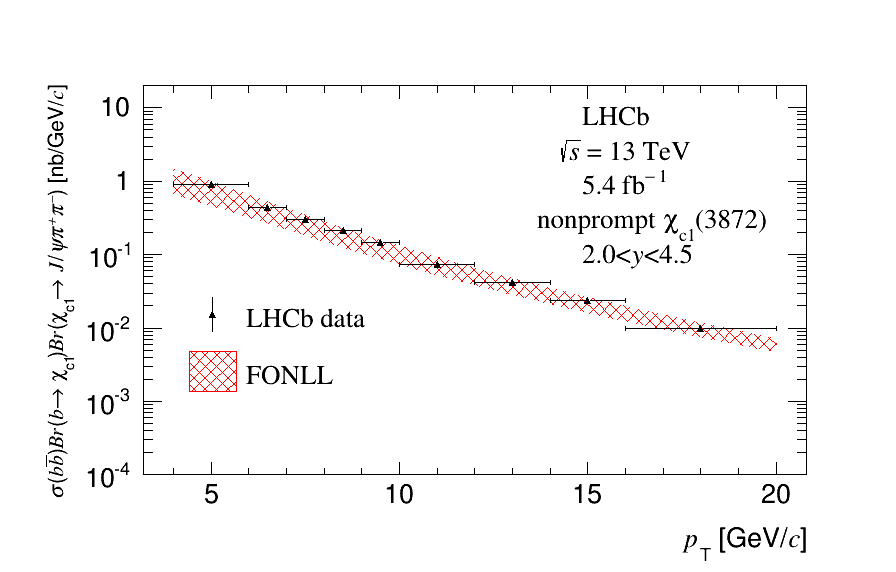} 
    \vspace*{-0.5cm}
  \end{center}
  \caption{
Measured production cross-section times branching fractions as a function of \pt for (left) prompt \theX mesons compared to NLO NRQCD
predictions~\cite{Meng:2013gga} and (right) from \bquark decays compared to FONLL predictions~\cite{Cacciari:1998it,Cacciari:2015fta}.}
  \label{fig:cs_abs_x}
\end{figure}

As defined in Eq.~(\ref{eq:cs_ratio_fun}), the absolute production cross-section of the \theX state times the branching fraction can be calculated using the measured cross-section ratio times
\mbox{$\sigma_{\psitwos}\BF(\psitoJpsipipi)$}.  The value of $\sigma_{\psitwos}$ is taken from the $\psitwos\to\mup\mun$ analysis~\cite{LHCb-PAPER-2018-049}. The world average for the \psitwos\to\jpsipipi branching fraction is $\BF(\psitoJpsipipi) = (34.68\pm 0.30) \times 10^{-2}$~\cite{PDG2020}. 
Figure~\ref{fig:cs_abs_x} shows the measured cross-section times branching fractions as a function of \pt for prompt \theX mesons compared to NLO NRQCD
predictions~\cite{Meng:2013gga} and from \bquark decays compared to FONLL predictions~\cite{Cacciari:1998it,Cacciari:2015fta}.
The prompt \theX production in NLO NRQCD can be expressed as 
  \begin{equation}
    \label{eq:cs_nrqcd}
d\sigma(pp\rightarrow\theX) = d\sigma(pp\rightarrow\chi_{c1}(2P))\cdot k,
  \end{equation}
where $k=Z_{c\bar{c}}\cdot\BF(\XtoJpsipipi)$, and $Z_{c\bar{c}}$ is the probability of the $\chi_{c1}(2P)$ component in the \theX. A fit was performed to the CMS  data~\cite{Chatrchyan:2013cld} using Eq.~\ref{eq:cs_nrqcd}, 
and a value of $k=0.014\pm0.006$ is obtained~\cite{Meng:2013gga}.
The prompt production is consistent with NLO NRQCD in the $\pt>10\gevc$ region.
The same settings of FONLL as for \psitwos mesons are used, except that
$\BF(b\rightarrow\theX)\BF(\theX\to\jpsi\pip\pim) = (4.3\pm 0.5) \times 10^{-5}$ is taken from this analysis. The FONLL calculation is also consistent with the measurement.
The absolute production cross-section can be derived from this result by using the recently measured $\BF(\XtoJpsipipi)=(4.1\pm1.3)\%$ \cite{Lees:2019xea} but the precision is insufficient to further improve the comparison with the various predictions.

\clearpage
\clearpage
\addcontentsline{toc}{section}{References}
\bibliographystyle{LHCb}
\bibliography{main,standard,LHCb-PAPER,LHCb-CONF,LHCb-DP,LHCb-TDR}
 
\newpage
% LHCb collaboration author list
% Data extracted on September 10th, 2021 at 4:49pm for paper reference LHCb-PAPER-2021-026
\centerline
{\large\bf LHCb collaboration}
\begin
{flushleft}
\small
R.~Aaij$^{32}$,
A.S.W.~Abdelmotteleb$^{56}$,
C.~Abell{\'a}n~Beteta$^{50}$,
F.J.~Abudinen~Gallego$^{56}$,
T.~Ackernley$^{60}$,
B.~Adeva$^{46}$,
M.~Adinolfi$^{54}$,
H.~Afsharnia$^{9}$,
C.~Agapopoulou$^{13}$,
C.A.~Aidala$^{87}$
% ...

{\footnotesize \it

% ...
$^{9}$Universit{\'e} Clermont Auvergne, CNRS/IN2P3, LPC, Clermont-Ferrand, France\\
% ...
$^{13}$LPNHE, Sorbonne Universit{\'e}, Paris Diderot Sorbonne Paris Cit{\'e}, CNRS/IN2P3, Paris, France\\
% ...
$^{32}$Nikhef National Institute for Subatomic Physics, Amsterdam, Netherlands\\
% ...
$^{46}$Instituto Galego de F{\'\i}sica de Altas Enerx{\'\i}as (IGFAE), Universidade de Santiago de Compostela, Santiago de Compostela, Spain\\
% ...
$^{50}$Physik-Institut, Universit{\"a}t Z{\"u}rich, Z{\"u}rich, Switzerland\\
% ...
$^{54}$H.H. Wills Physics Laboratory, University of Bristol, Bristol, United Kingdom\\
% ...
$^{56}$Department of Physics, University of Warwick, Coventry, United Kingdom\\
% ...
$^{60}$Oliver Lodge Laboratory, University of Liverpool, Liverpool, United Kingdom\\
% ...
$^{68}$Syracuse University, Syracuse, NY, United States\\
% ...
$^{87}$University of Michigan, Ann Arbor, United States, associated to $^{68}$\\
\bigskip
% ...
\medskip
}
\end{flushleft}

\end{document}